\documentclass[journal]{IEEEtran}
\usepackage{graphicx}
\graphicspath{ {images/} }
\usepackage{geometry}
\usepackage{mathtools,amsmath,amssymb,systeme,braket}
\usepackage{hyperref}
\usepackage{ragged2e}
\usepackage{booktabs}
\usepackage{float}
\usepackage[font=footnotesize]{caption}
\usepackage{caption}
\usepackage{placeins}
\usepackage[table,xcdraw]{xcolor}
\usepackage{authblk}
\usepackage[bottom]{footmisc}
\usepackage[utf8]{inputenc}
\usepackage[catalan,english]{babel}
\usepackage{subcaption}
\usepackage{colortbl}
\usepackage{titling}
\usepackage{makeidx}
\usepackage{fancyhdr}
\usepackage[table]{xcolor}
\usepackage{cite}

\definecolor{lightergray}{RGB}{245,245,245}
\definecolor{lightblue}{RGB}{220,230,245}

\makeindex

\newcommand{\subtitle}[1]{%
  \posttitle{%
    \par\end{center}
    \begin{center}\large#1\end{center}
    \vskip0.5em}%
}

\newcommand\lil {\hspace{1mm}}

\usepackage[labelfont=bf]{caption}

\title{An Analytical Approach to Design Space Exploration for Cavity-Mediated Quantum State Transfer in Multi-core Architectures}
\author{
Biel Pons Zaragozà$^{\ast}$, Junaid Khan$^{\dagger}$, Rohit Sarma Sarkar$^{\dagger}$, Sahar Ben Rached$^{\dagger}$, \\ Carmen G. Almudéver$^{\S}$, Eduard Alarcón$^{\dagger}$, and Sergi Abadal$^{\dagger}$}
\affil {$^{\ast}$\textit{Universitat de Barcelona, Spain} \\
$^{\dagger}$\textit{Universitat Politècnica de Catalunya, Spain} \\
$^{\S}$\textit{Universitat Politècnica de València, Spain}}

\begin{document}
\date{}
\maketitle


\selectlanguage{english}
\begin{abstract}

In multi-core quantum computing architectures, waveguide-mediated interconnects are essential for facilitating fast, high-fidelity quantum state transfer between qubits located in different chips. However, optimizing these systems typically relies on computationally expensive numerical simulations that offer limited physical insight. In this work, we derive exact analytical expressions for the state transfer dynamics of a two-qubit system coupled via a waveguide, modeled through a Jaynes-Cummings Hamiltonian and the Lindblad master equation. We apply the Monte Carlo wave-function method and obtain a closed-form solution for qubit occupation probabilities that accounts for both detuning and dissipative losses. Our analytical framework provides a significant computational speedup compared to standard numerical solvers, enabling large-scale parameter sweeps while maintaining high precision in both fidelity and latency predictions. Furthermore, the model reveals and explains systematic low-fidelity regions arising from destructive interference between internal oscillations and detuning-induced envelopes, which are phenomena that are difficult to characterize through numerical means alone. Finally, we propose a simplified latency model and an efficiency-based function to enable rapid identification of optimal operating points. This analytical approach provides a robust foundation for the design and optimization of interconnects in multi-core quantum processors.

\end{abstract}

\begin{IEEEkeywords}
Quantum computing, waveguide QED, latency, fidelity, detuning, Hamiltonian, Hermiticity, state transfer, Fock state, occupation number.
\end{IEEEkeywords}
\section{Introduction}

The advancement of quantum computing technology promises the realization of sophisticated calculations that remain fundamentally inaccessible to classical computers \cite{preskill2018quantum}. However, as the field transitions toward large-scale processors, a primary bottleneck emerges viz. the integration of a vast number of qubits within a single monolithic chip. This dense integration introduces significant engineering challenges, most notably the emergence of inter-qubit crosstalk and the degrading effects of qubit decoherence, both of which severely degrade overall system performance \cite{jnane2022multicore, niu2024multi}.

To circumvent these scaling limits, modular quantum architectures have emerged as a leading paradigm. In these systems, smaller, manageable quantum cores are linked via quantum-coherent interconnects \cite{rached2025characterizing, ang2024arquin}. The viability of such multi-core architectures depends fundamentally on the quality of the quantum state transfer across chip-to-chip links \cite{niu2023low}. Specifically, the system must achieve high-fidelity quantum state transfer while operating within a latency budget compatible with the coherence times required for circuit execution \cite{escofet2023interconnect}.

Several quantum communication channel technologies have been proposed, ranging from teleportation-based protocols using entangled photon pairs to cavity-mediated couplings between adjacent chips \cite{llewellyn2020chip, zhong2021deterministic, magnard2020microwave}. Among these, waveguide-based channels offer a particularly promising approach, acting as a quantum bus that confines electromagnetic modes to enhance light-matter interaction between spatially separated qubits \cite{blais2021circuit}. This enables deterministic quantum state transfer while balancing the high speed of direct coupling with the distance capabilities of fiber-based networks.  

Integrating such channels, however, introduces qubit decay and waveguide photon loss as competing sources of degradation that jointly shape system performance. Understanding how coupling strength $g$, waveguide-qubit detuning $\Delta_\omega$, qubit decay rate $\gamma$, and waveguide decay rate $\kappa$ govern fidelity and latency requires either exhaustive numerical simulation or an analytical model that makes these dependencies explicit. Exploring an extensive parameter space purely numerically is computationally demanding and provides limited insight into parameter dependencies \cite{khan2025, rached2024benchmarking}.
\begin{figure}
    \centering
    \includegraphics[width=\linewidth]{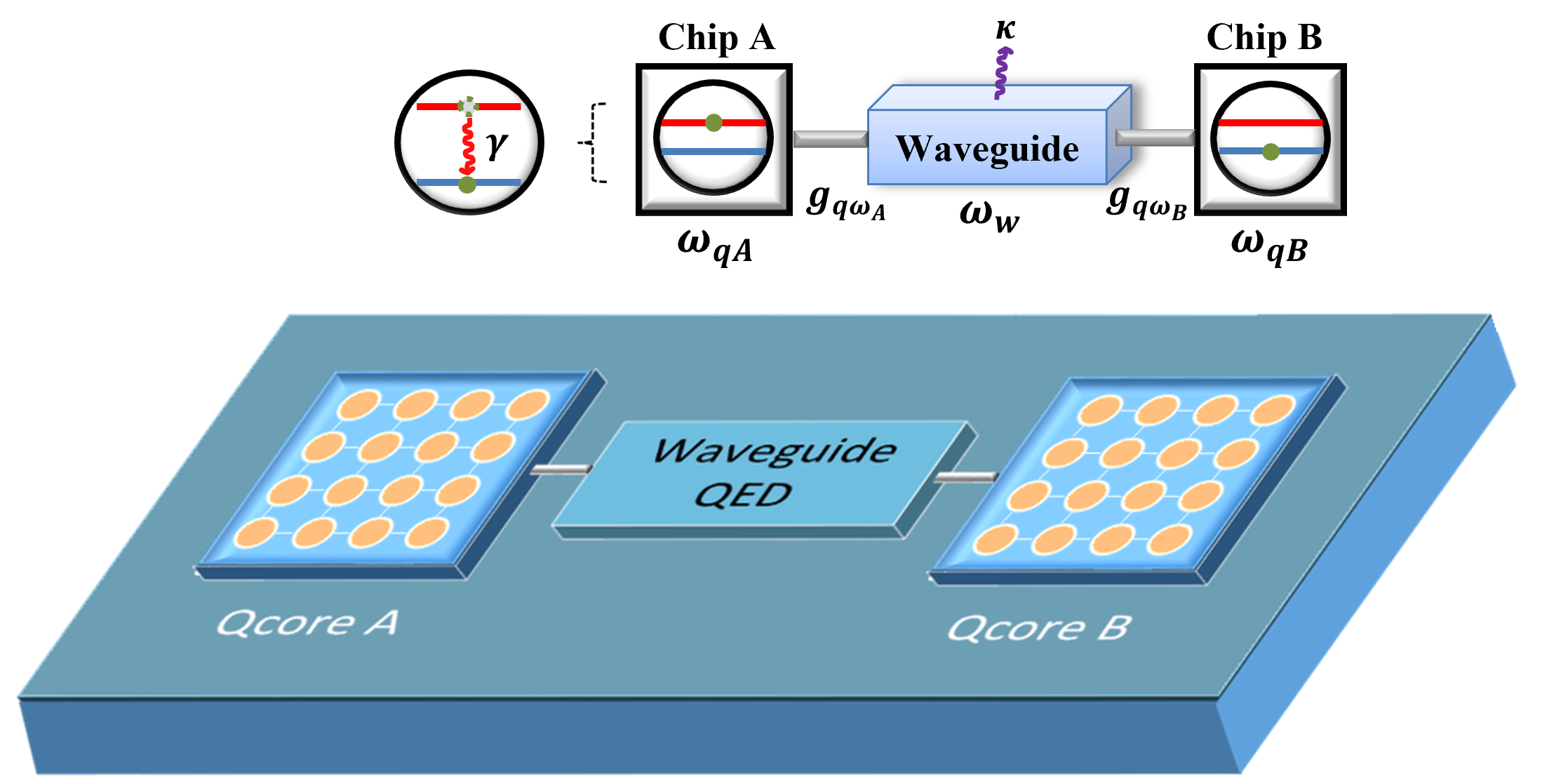}
    \caption{Schematic of a waveguide-mediated interconnect 
between two qubit chips in a modular 
quantum computing system.}
    \label{fig: System Diagram}
\end{figure}
In this work, we derive closed-form analytical expressions for the state occupation probabilities of a waveguide-coupled two-qubit system under both lossless and lossy conditions. These expressions are two orders of magnitude faster to evaluate than QuTiP simulations \cite{johansson2012qutip} and directly expose the influence of each system parameter on performance, enabling rapid parameter space exploration and optimization without sacrificing physical insight. The approach is based on the \textit{Monte Carlo wave-function} method \cite{molmer1993monte}, in which quantum jumps can be neglected for this system, yielding a fully analytical time evolution without time-step discretization.

Despite this speedup, the model produces results with high precision compared to the numerical simulations. This depicts that the analytical model is advantageous and has the potential to be integrated into automated design automation \cite{edaq2025, almudever2024} tools for real-time optimization of large-scale quantum networks. By replacing slow numerical solvers with exact closed-form expressions, researchers can perform instantaneous sensitivity analyses and explore vast parameter spaces that were previously inaccessible.

The remainder of this paper is organized as follows. In Section \ref{Related work}, we review related work in modular quantum architectures and analytical methods. In Section \ref{Prelimanries}, we go over the preliminaries, including the system Hamiltonian and the Lindblad master equation framework. In Section \ref{Analytical 4}, we derive the exact analytical expressions for both lossless and lossy systems using the Monte Carlo wave-function method. In Section \ref{Simulation}, we validate these expressions against numerical simulations, quantifying the computational speedup. In Section \ref{Analysis}, we provide a detailed analysis of the fidelity regions, including the identification of systematic low-fidelity zones and the proposal of a latency-fidelity optimization model. Finally, we conclude the paper and discuss potential future extensions in Section \ref{Conclusion}.



\section{Related work}
\label{Related work}
Scaling quantum processors beyond the limits of monolithic chip designs has driven significant interest in modular architectures, where multiple quantum processors units are linked through quantum-coherent interconnects \cite{bravyi2022future, 
escofet2023interconnect}. The core challenge in such systems is not merely connecting chips physically, but preserving quantum coherence while keeping transfer latency within the bounds of a practical circuit execution. 

Several physical mechanisms have been explored for the inter-chip quantum state transfer. Direct capacitive coupling between superconducting chips using tunable couplers has been demonstrated at the two-qubit gate level, with fidelities competitive with on-chip operations~\cite{field2023}. However, such approaches are inherently short-range and require precise physical proximity between chips. Waveguide and cavity-mediated channels offer an alternative that relaxes this spatial constraint \cite{leung2019deterministic}. A comparative benchmarking of cavity-mediated interconnect technologies across different decay-rate and coupling-strength configurations shows that only specific parameter regions meet the fidelity and latency thresholds required for reliable chip-to-chip links ~\cite{rached2024benchmarking}. Numerical studies of the waveguide-coupled two-qubit scenario using QuTiP simulations have further characterized how detuning, coupling strength, and loss rates jointly determine performance ~\cite{khan2025}. 

A parallel line of research has addressed quantum state transfer in open systems using analytical methods. For weakly coupled spin-chain channels, closed-form relations between transfer time and fidelity show that homogeneous chains offer the best speed-fidelity tradeoff~\cite{cosme2018}. In waveguide-linked optomechanical systems, analytical fidelity formulas obtained from a scattering-Lindblad framework show that suitable control trajectories can exceed 96\% fidelity even when losses are asymmetric~\cite{soh2021high}. These studies consistently provide an inference that analytical models make parameter dependence explicit while avoiding the cost of repeated numerical simulations. At the hardware level, studies of inter-core qubit traffic in modular quantum processors identify communication overhead as a major limit on circuit performance, which makes the latency of each transfer event an important design variable ~\cite{rodrigo2023}. Despite this progress, no closed-form treatment has been reported for a waveguide-coupled two-qubit system that simultaneously captures occupation probability, transfer fidelity, and latency in the presence of both qubit and waveguide losses. The present work addresses that gap directly through the mathematical derivation of the state transfer equation and a model to optimize the fidelity-latency tradeoff. 


\section{Prelimanries}
\label{Prelimanries}

In this section, we provide the groundwork for the closed form of the quantum state transfer using a waveguide channel. First we discuss about some preliminary tools required for the proposed analytical model. The quantum system considered in this work is multicore in nature and consists of two qubit chips (A and B) coupled through a waveguide, Fig.~\ref{fig: System Diagram}. For the sake of simplicity and understanding the qubit-waveguide model in depth, we follow the work done by Khan et al. \cite{khan2025} and provide an analytical characterization of the same. Hence, we have considered only one mode of the waveguide that governs the different energies of the photon transferred. An analysis with the same approach for a large number of waveguides is left for future work.

The standard way to model qubit-waveguide interactions is to treat them as Jaynes-Cummings-type under the \textit{Rotating Wave Approximation (RWA)} ~\cite{Azem2016JaynesCummingsM}. This is an energy-conserving exchange of quanta where the qubit absorbs a photon and gets excited or emits a photon when reaching ground state ~\cite{Azem2016JaynesCummingsM}. The dynamics of this proposed model are governed by a Hamiltonian operator as expressed in Eq.~\eqref{Hamiltonian} obtained as the sum of the individual qubit and waveguide Hamiltonians (local Hamiltonians), together with their Jaynes--Cummings--type interaction \cite{reiserer2015cavity} \footnote{For simplicity, $\hbar = 1$ will be considered throughout the paper.},
\begin{equation}\label{Hamiltonian}
    \begin{aligned}
H ={}& \omega_q \sigma_A^+ \sigma_A^-
      + \omega_q \sigma_B^+ \sigma_B^-
      + \omega_{WG} a^\dagger a \\
    &+ g \left(
      \sigma_A^+ a + \sigma_A^- a^\dagger
      + \sigma_B^+ a + \sigma_B^- a^\dagger
      \right),
\end{aligned}
\end{equation}
where $\omega_q$ is the resonance frequency of both qubits, $\omega_{WG}$ is the resonance frequency of the waveguide and $g$ is the coupling strength between the qubits and the waveguide. 
$\sigma^\pm_{A}, \sigma^\pm_{B}$ are the raising and lowering operators acting on qubits A and B, respectively. Similarly, $a$ and $a^\dagger$ are the annihilation and creation operators for photons acting on the waveguide. If the system was closed, the state transfer dynamics would have been described by the Time-dependent Schr\"{o}dinger equation~\cite{Schrodinger}, $i \hbar \frac{d}{dt} \lvert \Psi (t) \rangle=\hat{H} \lvert \Psi (t) \rangle$. However, for open quantum system dynamics i.e. where the system interacts with the environment, dissipative processes need to be incorporated. In this regard, the Lindblad master as shown in Eq~\eqref{Lindblad} provides a general description of the open-system dynamics with dissipation and decoherence \cite{manzano2020short}. The equation is as follows,
\begin{equation}\label{Lindblad}
\frac{d\rho}{dt} = -i[H, \rho] + \sum_{k=1}^3 \left( L_k \rho L_k^{\dagger} - \frac{1}{2} \{ L_k^{\dagger} L_k, \rho \} \right),
\end{equation}
where, $L_k$'s are the collapse operators acting on each of the three components: $L_1= \sqrt{\gamma} \sigma^-_A$ for qubit A, $L_2=\sqrt{\gamma} \sigma^-_B$ for qubit B, and $L_3= \sqrt{\kappa} a$ for the waveguide. These operators act as losses in the proposed model that affect the state transmission. The metrics for assessing the quality of such transmissions are fidelity and latency. Given two density matrices of quantum states $\rho$ and $\sigma$, the fidelity \cite{Jozsa01121994} between them is defined to be the following 
\begin{equation}
    \label{eq:fidelity}
    F(\rho, \sigma) = \left ( \text{tr} \sqrt{ \sqrt{\rho} \sigma \sqrt{\rho}}\right)^2
\end{equation}
In our case, $\rho$ corresponds to the initial state of qubit A and $\sigmaº$ to the state of qubit B at time $t$. A fidelity of
$F=1$ indicates that qubit B has perfectly acquired the original state of qubit A. Since fidelity evolves over time, we define the transfer fidelity for a given set of parameters as the maximum value of $F$ achieved throughout the evolution. Complementarily, the latency \cite{stute2012tunable} characterizes the temporal efficiency of the transfer, and is defined as the time at which the fidelity at qubit B reaches its prominent peak. Together, the peak fidelity and the latency at which it occurs provide a complete characterization of the state transfer quality. However, a high peak fidelity does not necessarily imply an efficient transfer, as the maximum may occur at impractically long times. This tradeoff between fidelity and latency is further discussed in Annex~\ref{sec:efficiency}.\\


Within this framework, the system's evolution is fully described by the Hamiltonian mentioned in  Eq.~\eqref{Hamiltonian} and the Lindblad in Eq~\eqref{Lindblad}. However, this results in a quantum Langevin equation~\cite{Cirac_1997} for the density operator. This description can be further simplified into a time-dependent Schr\"{o}dinger equation using an effective Hamiltonian instead of the original one. In the following, we discuss the methodology to solve the equations for the lossless system. Subsequently, we develop the equations for the lossy system as well.

\section{Analytical Model for Cavity-Mediated Quantum State Transfer}
\label{Analytical 4}
In this section, we explained the method for deriving the time-dependent probabilities of finding the qubit on Chip B in the excited state, following the transfer of a quantum state from the qubit on Chip A through the waveguide. The Hilbert space of the system is given by the tensor product of the independent Hilbert spaces of its components,
\begin{equation}
    \label{eq:Hilbert_space}
    \mathcal{H} = \mathcal{H}_{\text{Qubit A}} \otimes \mathcal{H}_{\text{Waveguide}} \otimes \mathcal{H}_{\text{Qubit B}}.
\end{equation}

\begin{figure*}[t]
    \centering
    \begin{subfigure}{0.35\linewidth}
        \centering
        \includegraphics[width=\linewidth]{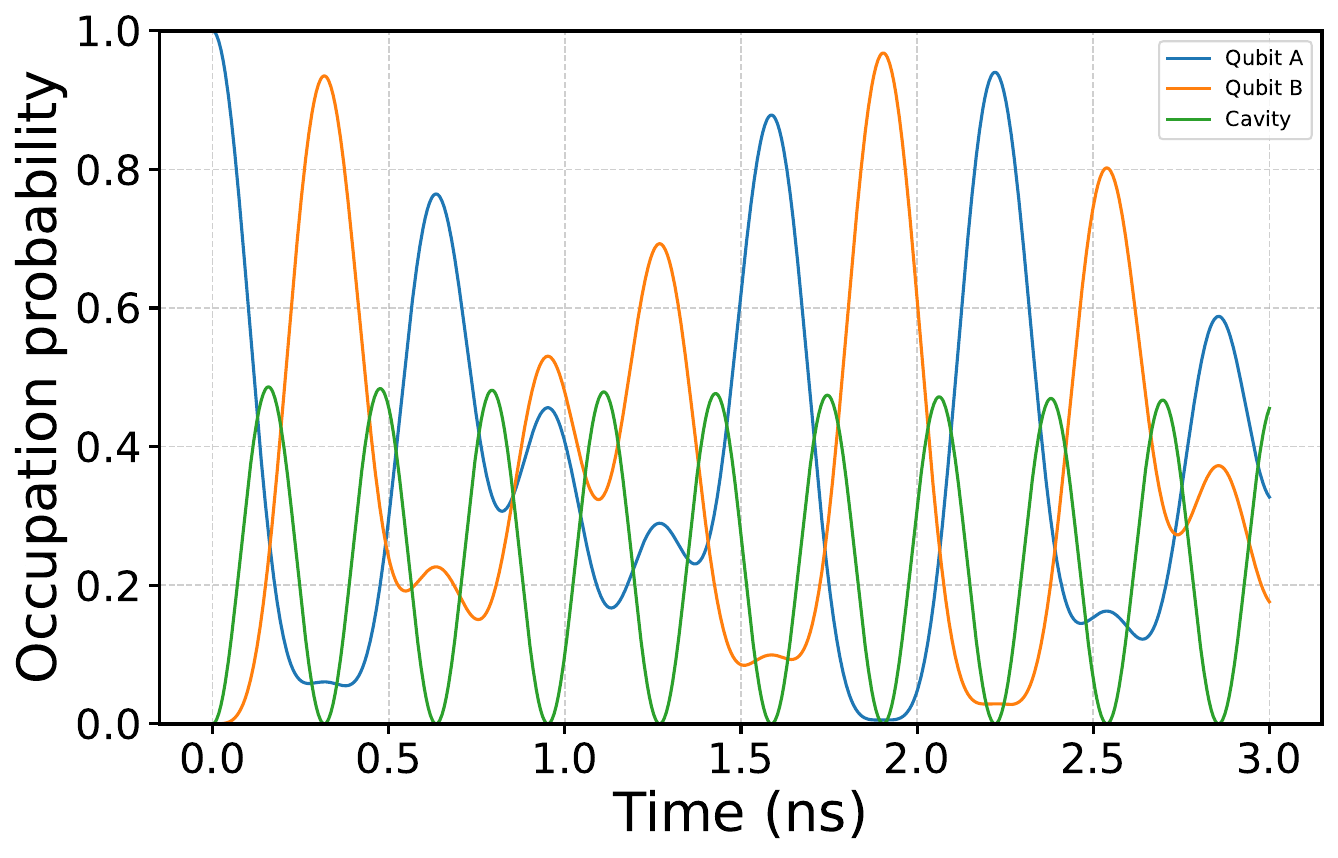}
        \caption{Analytical Simulation}
        \label{fig:analytical}
    \end{subfigure}
    \begin{subfigure}{0.35\linewidth}
        \centering
        \includegraphics[width=\linewidth]{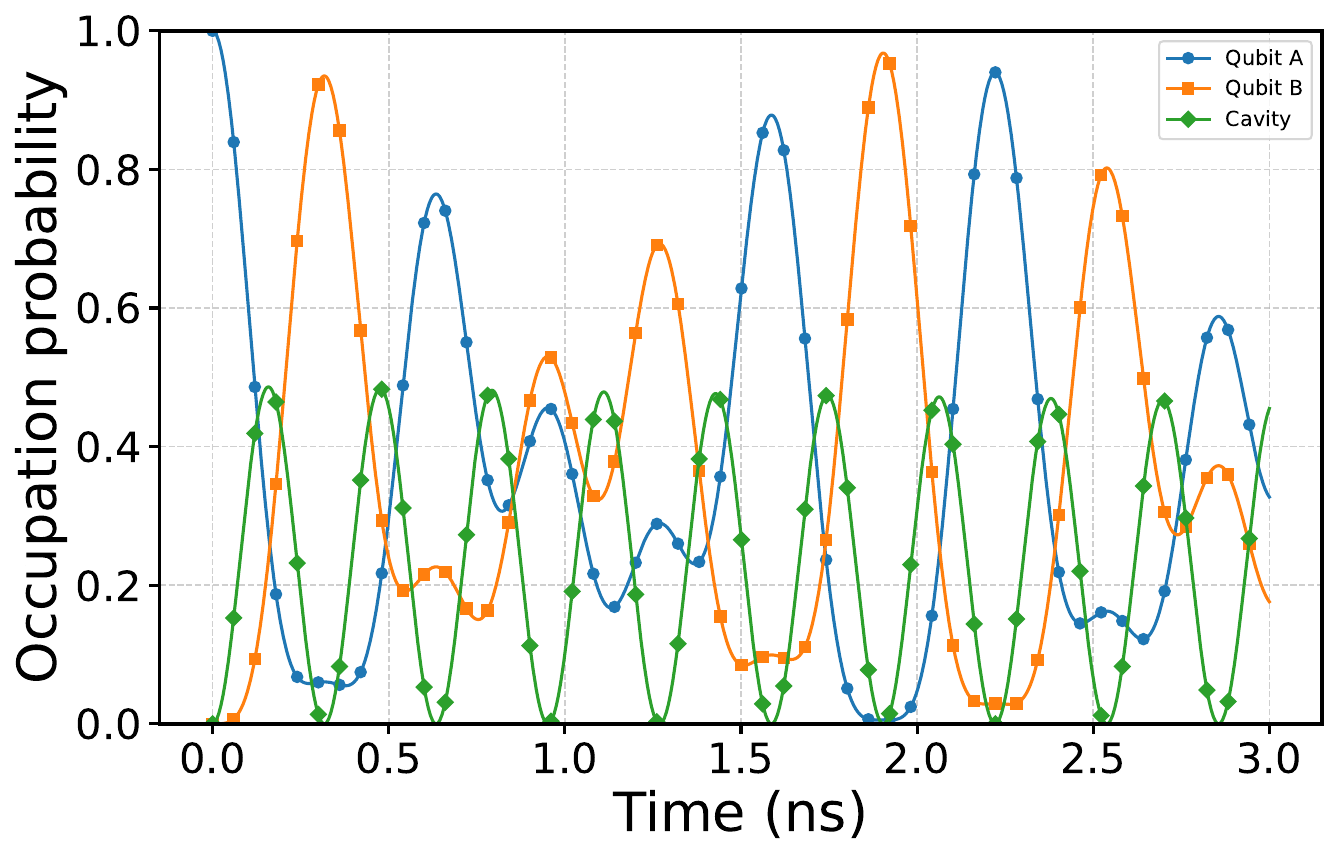}
        \caption{QuTip Simulation}
        \label{fig:simulation}
    \end{subfigure}
    \caption{Comparison between plots using the analytical equation and using QuTip for $\Delta_{\omega} = 500$ MHz, $g = 1.1$ GHz, $\kappa = 3$ MHz and $\gamma = 2$ MHz.}
    \label{probability comparison}
\end{figure*}

Where the system is initially in the Fock state \cite{Dirac_1927} $|1,0,0\rangle$, with just qubit A excited, and the target state is $|0,0,1\rangle$, with just qubit B excited. 
The probabilities of finding qubit B in the excited state are characterized by their corresponding evolution operators, which we shall discuss in this section as well.  
\subsection{Lossless system}
\label{Subsection:lossless}
Let us first consider a system without losses with $\kappa = \gamma = 0$. The time evolution of the occupation number \cite{Dirac_1927} for each component can be found simply using the unitary time evolution operator $U(t) = e^{-iHt}$. 
Since the waveguide Hilbert space is infinite-dimensional, the total Hilbert space in Equation~\eqref{eq:Hilbert_space} is likewise infinite-dimensional. Consequently, working directly with the Hamiltonian in this space can be challenging. But under the \textit{RWA} and with just one wave mode of the waveguide, the only Fock states \cite{Dirac_1927} of the waveguide that contribute to the dynamics, are $\lvert 0 \rangle$ and $\lvert 1 \rangle$ (demonstrated in the Appendix \ref{single_excitation}). Since each qubit is a two-level system, the full Hilbert space is eight-dimensional ($2^3$). However, \textit{RWA} restricts the dynamics to the subspace spanned by $\{\,|1,0,0\rangle, |0,1,0\rangle, |0,0,1\rangle\,\}$. For this reason, the Hamiltonian can be simply expressed as a $3 \times 3$ matrix, 
\begin{equation}
\label{eq:Lossless_Hamiltonian}
    H =
\begin{pmatrix}
\omega_A & g & 0 \\
g & \omega_{WG} & g \\
0 & g & \omega_B
\end{pmatrix}
\end{equation}
A straightforward calculation, as depicted in Appendix \ref{Lossless_equations}, shows that the probability of finding the qubit B on the excited state at time $t$ is the following,
\begin{equation}\label{Lossless_Prob}
\begin{aligned}
P_B(t) = {}& \frac{1}{4} \Bigl(
  1 + \cos^2(\theta t)
  - 2 \cos(\theta t)\cos(\delta t) \\
&\quad
  - \frac{2\delta}{\theta} \sin(\theta t)\sin(\delta t)
  + \frac{\delta^2}{\theta^2} \sin^2(\theta t)
\Bigr)
\end{aligned}
\end{equation}

\[
\delta = \Delta_{\omega} / 2 \qquad 
\Omega = \sqrt{2} g 
\qquad 
\theta = \sqrt{\Omega^2 + \delta^2}
\]
where ideal transmission will be considered when $P_B = 1$. Equation \eqref{Lossless_Prob} is derived directly by extracting the eigen-spectrum of the Hamiltonian operator. Notice that the oscillation frequencies of the occupation probability ($\theta$ and $\delta$) are not directly dependent on the frequencies of either the qubits or the waveguide, but rather by the detuning between them ($\Delta_{\omega} = \omega_\text{WG} - \omega_q$) and the coupling strength ($g$). This analytical expression is derived as a precursor to a more complex system where losses are introduced.

\subsection{Lossy System}
Here, we work with the system with losses, $\kappa \neq 0$ and $\gamma \neq 0$, by adapting a similar approach to Section \ref{Subsection:lossless}. We integrate the dissipation and decoherence dynamics of the Lindblad master equation directly into the Hamiltonian via the Monte Carlo Wave Function method \cite{Castin}. Afterwards, each state evolves following a Schr\"{o}dinger equation with the new non-Hermitian Hamiltonian. The Monte Carlo Wave Function method consists of the evolution of a non-Hermitian Hamiltonian and randomly decided quantum jumps (which do not affect our system, as mentioned in the Appendix \ref{Appendix:Jumps}). Mathematically, it is a way of incorporating the anti-commutator term into the Hamiltonian, creating a new matrix that we can consider as an \textit{``Effective Hamiltonian"}. In other words, this new Hamiltonian must satisfy the following condition
\begin{equation}
    -i\!\left(H_{\mathrm{eff}}\rho - \rho H_{\mathrm{eff}}^{\dagger}\right)
= -i[H,\rho]
- \sum_{k=1}^3 \tfrac{1}{2} \{ L_k^{\dagger} L_k, \rho \}
\end{equation}
Rearranging the terms:
\begin{equation}
\begin{aligned}
    H_{\mathrm{eff}}\rho - \rho H_{\mathrm{eff}}^{\dagger} = {} & \left( H - \frac{i}{2} \sum_{k=1}^3  L_k^{\dagger} L_k \right)\rho \\ - &\rho\left( H + \frac{i}{2} \sum_{k=1}^3  L_k^{\dagger} L_k \right)
\end{aligned}
\end{equation}
From here, we can observe that the non-Hermitian Hamiltonian takes the following form:
\begin{equation}
    H_{\text{eff}} = H - \frac{i}{2} \sum_{k=1}^3  L_k^{\dagger} L_k
\end{equation}
That can be once again expressed in the single-excitation subspace spanned by the basis vectors $\{\,|1,0,0\rangle, |0,1,0\rangle, |0,0,1\rangle\,\}$. 
\begin{equation}
    H_{\text{eff}} = 
\begin{pmatrix}
\omega_q - i\frac{\gamma}{2} & g & 0 \\
g & \omega_{WG} - i\frac{\kappa}{2} & g \\
0 & g & \omega_q - i\frac{\gamma}{2}
\end{pmatrix}
\end{equation}
And now, the time evolution for each component can be found using the non-unitary time evolution operator $e^{-iH_\text{eff}  t}$. The full mathematical derivation is explained in detail in the Appendix \ref{Appendix:Lossy} of this paper. The main idea is to merge the Lindblad term into the Hamiltonian in order to treat it in a similar manner as the lossless system. This results in the following equation for the probability of excitation of the qubit B at time $t$.
\begin{equation}\label{Lossy_Prob}
\begin{aligned}
        P_B(t) ={}& \frac{1}{4} \Bigl(
e^{-\gamma t}
- 2 e^{-\frac{\kappa + 3\gamma}{4}t}
[\,A(t) \cos(\delta t) +\\ & B(t) \sin(\delta t)\,] + e^{-\frac{\kappa+\gamma}{2}t}(A^2(t) + B^2 (t))
\Bigr)
\end{aligned}
\end{equation}

With 
\begin{equation}
    \left\{
\begin{aligned}
A(t) ={}& \cos(\theta t)\,\cosh(\phi t) 
     - a\cos(\theta t)\,\sinh(\phi t) \\
     & - b\sin(\theta t)\,\cosh(\phi t) \\[2mm]
B (t) ={}& -\sin(\theta t)\,\sinh(\phi t) 
     + a\sin(\theta t)\,\cosh(\phi t)\\ 
     &- b\cos(\theta t)\,\sinh(\phi t)
\end{aligned}
\right.\\
\end{equation}

It is of note that the parameters of the Equation~\eqref{Lossy_Prob}  changes with respect to the lossless system mentioned in Equation \eqref{Lossless_Prob}. This is due to the complex contributions of the collapse operators:
\begin{equation}
\begin{aligned}
\theta' = \sqrt{\Omega^2 + \delta^2 - \Gamma^2 + i\,2\delta\Gamma}
\qquad
\Gamma = \frac{\gamma - \kappa}{4}\\
\theta = \text{Re}{(\theta')}
\qquad
\phi = \text{Im}{(\theta')}
\qquad
a+ib ={} \frac{\delta + i\Gamma}{\theta'}
\end{aligned}
\end{equation}
For this case, the oscillation frequency of the occupation probabilities depends also on the difference between the qubit decay rate ($\gamma$) and waveguide decay rate ($\kappa$), apart from the detuning ($\Delta_\omega$) and coupling strength ($g$). It is important to emphasize that, in the regime of weak losses ($\kappa, \gamma \rightarrow 0$), the system behaves similarly to the lossless case, with the dynamics simply acquiring exponential decay, as expected. Once we know the occupation probability of the desired qubit, the latency is derived simply as the time with maximum occupation probability. Because, since we are dealing with pure states, fidelity is equivalent to the occupation probability at such time \cite{Jozsa01121994}.

\section{Accuracy and computational efficiency of the model}
\label{Simulation}
In this section, we provide both analytical and numerical simulations of probability distributions arising from the quantum state transfer model proposed in this paper. All the simulations have been carried out on a system with Processor 13th Gen Intel(R) Core(TM) i7-1355U (1.70 GHz), RAM memory 16 GB, on Python 3.12.7.

The computational cost of each simulation is quantified by measuring time using Python’s \texttt{time.perf\_counter()}. For each method, the execution time corresponds to the average over ten independent runs (after an initial warm-up execution) of the time required to compute \(N_t = 1000\) time steps from $0$ to $3$ ns. 

\subsection{Validation for the occupation probability over time}
\label{subsection:validation_Analytical}
In order to evaluate both the computational cost and the validity of our analytical equations, we compare the occupation probability predicted by the analytical equations with the results obtained from full QuTiP simulations. 
We plot the time evolution of the occupation probability of each component in Fig.~\ref{probability comparison}, as obtained from both numerical simulations and the analytical model, demonstrating excellent agreement between the two approaches.

Under the previously mentioned conditions, the time required to generate the occupation probability is \(T_{\mathrm{QuTiP}} = 0.017 \pm 0.002\)~s for the QuTiP simulations. On the other hand, the analytical equation requires \(T_{\mathrm{equation}} = 0.0004 \pm 0.0002\)~s. Comparing the computational cost for a range of time steps in Table~\ref{tab:time dependance N}, we observe that the time required for the QuTip simulations is consistently around two orders of magnitude larger than the required for the analytical equation. This demonstrates the effectiveness of the analytical model for large time steps. Furthermore, to achieve optimal transmission, it is necessary to perform a parameter sweep over the region spanned by $\Delta_\omega$ and $g$. Such a sweep requires the consistent generation of results over a large number of time steps. In this context, the analytical model is particularly advantageous due to its computational efficiency and accuracy. 
 
\begin{table}[]
    \centering
    \caption{Computational cost of evaluating the occupation probability for $N_t$ time steps, for the QuTiP simulations and the analytical equation. Results are averaged over ten independent runs.}
     \rowcolors{2}{lightergray}{}
    \begin{tabular}{|c|c|c|}
    \hline
    \rowcolor{lightblue} 
        $N_t$ & $T_\text{QuTip} (ms)$ & $T_\text{equation} (ms)$\\
    \hline
         $10^1$& $3 \pm 1$ &  $0.04 \pm 0.02$\\
         $10^2$&  $4 \pm 0.8$ &  $0.035 \pm 0.008$\\  
         $10^3$&  $17 \pm 2$ &  $0.4 \pm 0.2$\\
         $10^4$&  $150 \pm 20$ &  $1.1 \pm 0.3$\\
         $10^5$&  $1430 \pm 40$ &  $8.1 \pm 0.7$\\
         $10^6$&  $14000 \pm 200$ &  $138 \pm 3$\\
    \hline
    \end{tabular}
    \label{tab:time dependance N}
\end{table}

\subsection{Validation for the parameter space}
To understand the dependence of latency and fidelity with our parameters, we  plot heatmaps for different combinations of losses. In this paper, the heatmaps are computed on a $200 \times 200$ grid by varying the coupling strength $g$ between $10$ and $100$~MHz (avoiding $g \sim 0$ to have transmission) and the detuning $\Delta_{\omega}$ between $0$ and $100$~MHz. As depicted in Fig.~\ref{fig:Fidelity Heatmap}, we get very similar results using the analytical expression and using QuTip simulations with an average latency difference of $3 \ \mathrm{ps}$ and an average fidelity difference of $4.5\times10^{-6}$. 

The benchmark now corresponds to the computation of a full $200 \times 200$ latency heatmap with $N_t = 400$ time steps spanning $0$ to $250$~ns.
The total execution time amounts to \(T_{\mathrm{QuTiP}} = 1260 \pm 30\)~s for the QuTiP simulations and to \(T_{\mathrm{equation}} = 174 \pm 3\)~s for the equation simulations. 
This further validates the results presented in Sec.~\ref{subsection:validation_Analytical}, where the parameters were fixed, while in the present section they are varied. 





\section{Analysis of the Cavity-Mediated Quantum State Transfer Model}
\label{Analysis}
In this section, we analyze the conditions for optimal state transfer. We first study how each loss term individually affects the transmission, then identify the distinct fidelity regions of the parameter space. Finally, we propose a simple model that predicts the fidelity and latency for arbitrary parameter combinations in the small-loss regime.

\subsection{On the importance of losses}
\begin{figure}[]
    \centering
    \begin{subfigure}{0.45\linewidth}
        \centering
        \includegraphics[width=\linewidth]{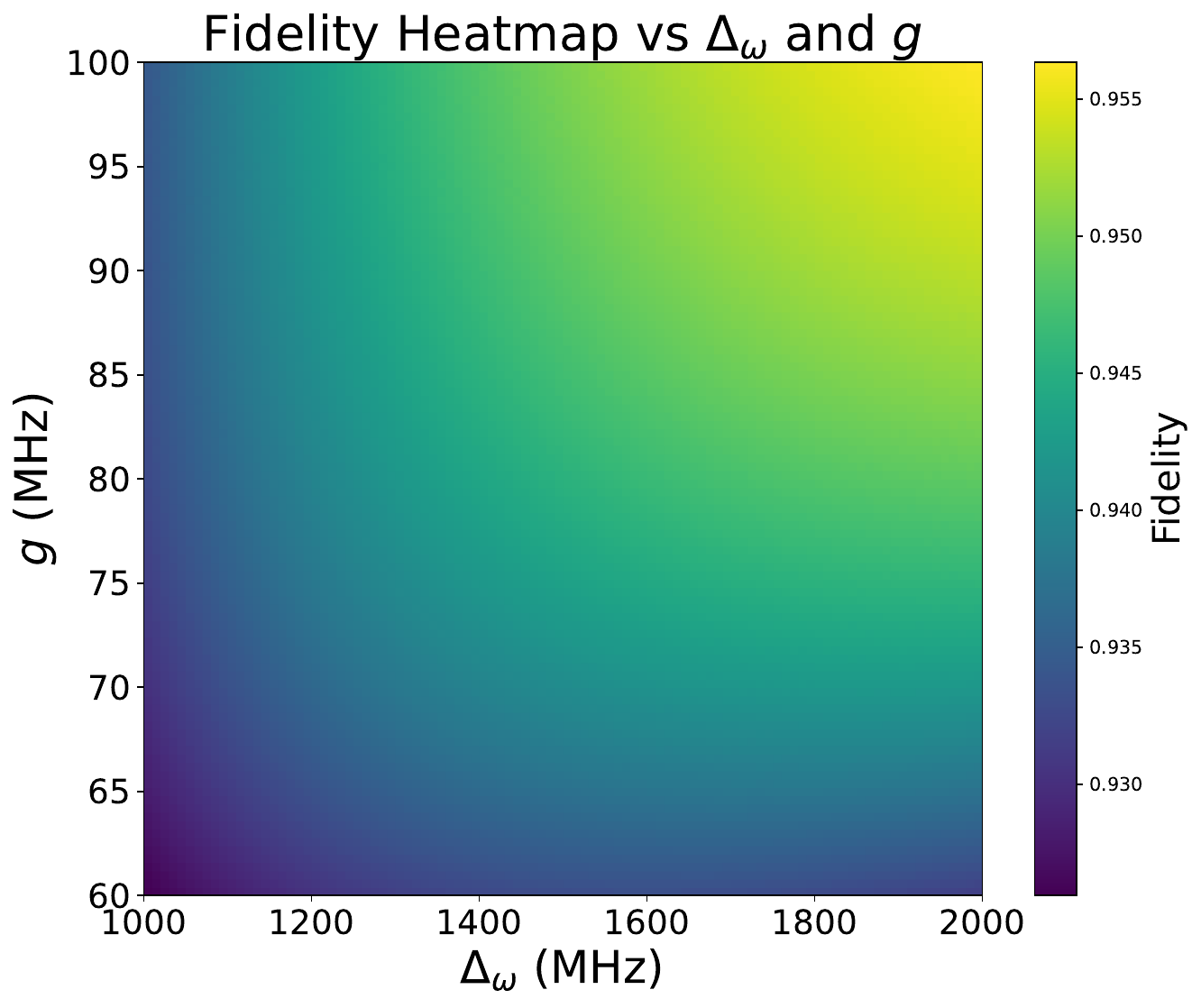}
        \caption{$\kappa = 70$ MHz and $\gamma = 0.1$ MHz.}
        \label{fig:kappa_loss}  
    \end{subfigure}
    \begin{subfigure}{0.45\linewidth}
        \centering
        \includegraphics[width=\linewidth]{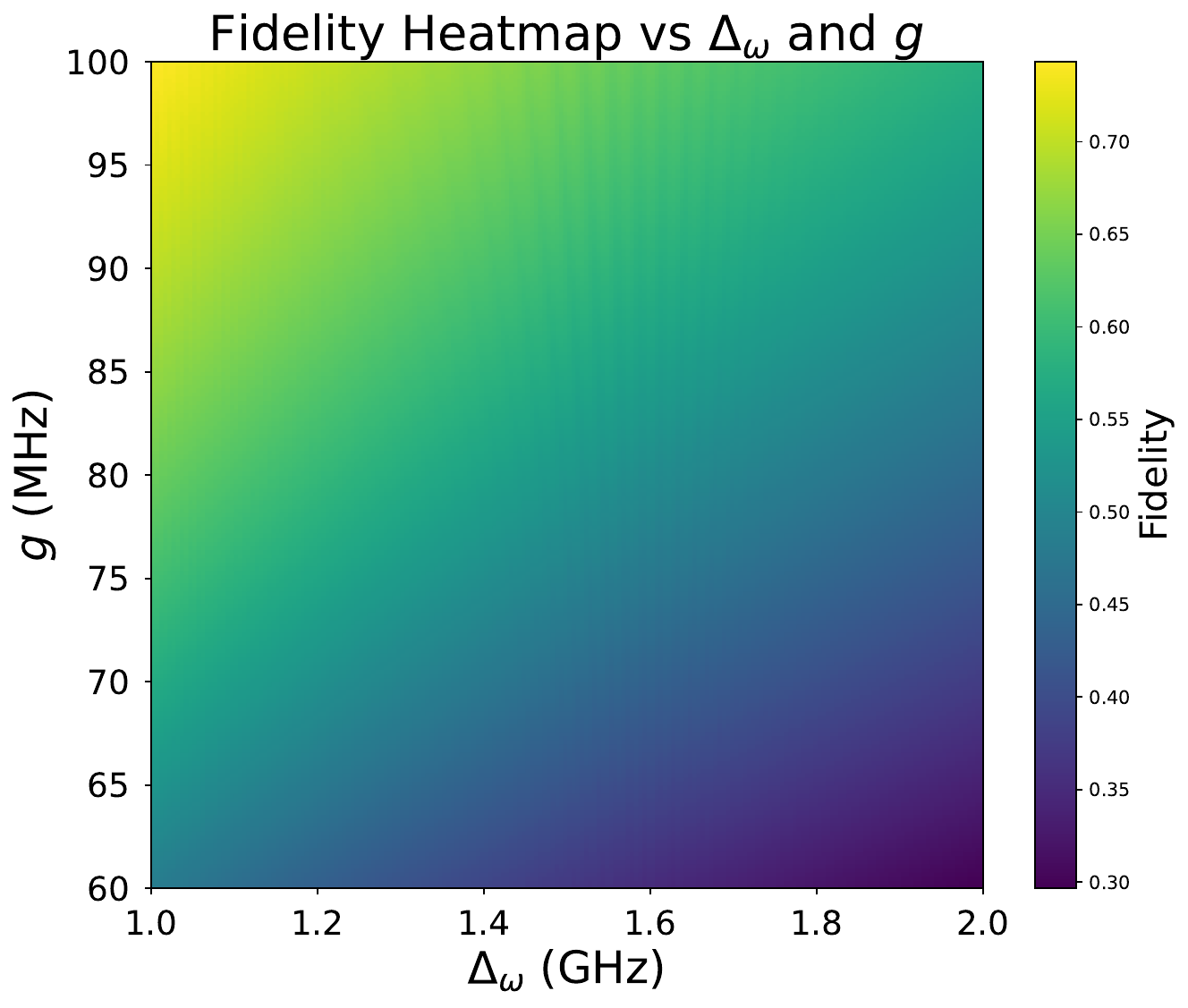}
        \caption{$\kappa = 0.1$ MHz and $\gamma = 4$ MHz.}
        \label{fig:gamma_loss}
    \end{subfigure}
    \caption{Transfer fidelity heatmap over coupling strength $g$ and detuning $\Delta_\omega$ for a highly detuned system over different combinations of losses.}
\end{figure}

It has been previously observed that, for some systems with high losses, it can be beneficial to detune the qubit frequency $\omega_q$ from the waveguide frequency $\omega_\text{WG}$ \cite{khan2025}, as shown in Fig.~\ref{fig:kappa_loss}. 
To study how each loss term affects the state transmission, we set the other one to zero in Equation~\eqref{Lossy_Prob}.
\subsubsection{Waveguide losses}
If we consider losses in the waveguide $\kappa \neq 0$ but no losses in the qubits, $\gamma = 0$, the occupation probability of qubit B is
        \begin{equation}
        \begin{aligned}
            P_B (t) ={}& \frac{1}{4} \Bigl( 1  +  e^{-\frac{\kappa}{2} t}(A^2(t) + B^2(t)) \\
            &- 2 e^{-\frac{\kappa}{4} t} [A(t)\cos(\delta t)+ B(t) \sin(\delta t)]\Bigr)
        \end{aligned}
        \end{equation}
Due to the exponential decay, for large $t$, fidelity will converge to $1/2$ for both qubits. Now that $\Gamma = - \frac{\kappa}{4}$, we get $\phi <0$ (in Appendix \ref{Appendix:Lossy} it is shown that $\text{sign}(\phi) = \text{sign}(\delta\Gamma) $) and $b<0$. Which changes the behavior of $A(t)$ and $B(t)$, making the term $A^2 + B^2$ increase with $\delta$. This means that detuning can improve fidelity when dealing with high waveguide losses $\kappa$, as seen in Figure~\ref{fig:kappa_loss}.\\

\subsubsection{Qubit losses}
If we consider losses in the qubits $\gamma \neq 0$ but no losses in the waveguide $\kappa = 0$, the occupation probability of the qubit B is,
        \begin{equation}
        \begin{aligned}
            P_B (t) = & \frac{1}{4} \Bigl ( e^{-\gamma t} + e^{-\frac{\gamma}{2} t}(A^2(t) + B^2(t)) \\  &- 2 e^{-\frac{3}{4} \gamma t} [A(t)\cos(\delta t)+ B(t) \sin(\delta t)]\Bigr )
        \end{aligned}
        \end{equation}
In this case, the whole system will be emptied for large $t$. Now that $\Gamma =  \frac{\gamma}{4}$, we get $\phi >0$ and $b>0$, and the behavior of $A(t)$ and $B(t)$ does not change from the original equation. This shows that detuning worsens fidelity when dealing with high qubit losses $\gamma$, as seen in Figure~\ref{fig:gamma_loss}. Hence, this demonstrates that detuning poses to be an important factor in state-transfer.

\subsection{Low fidelity regions}
\begin{figure*}[t]
    \centering
    \begin{subfigure}[t]{0.23\linewidth}
        \centering
        \includegraphics[width=\linewidth]{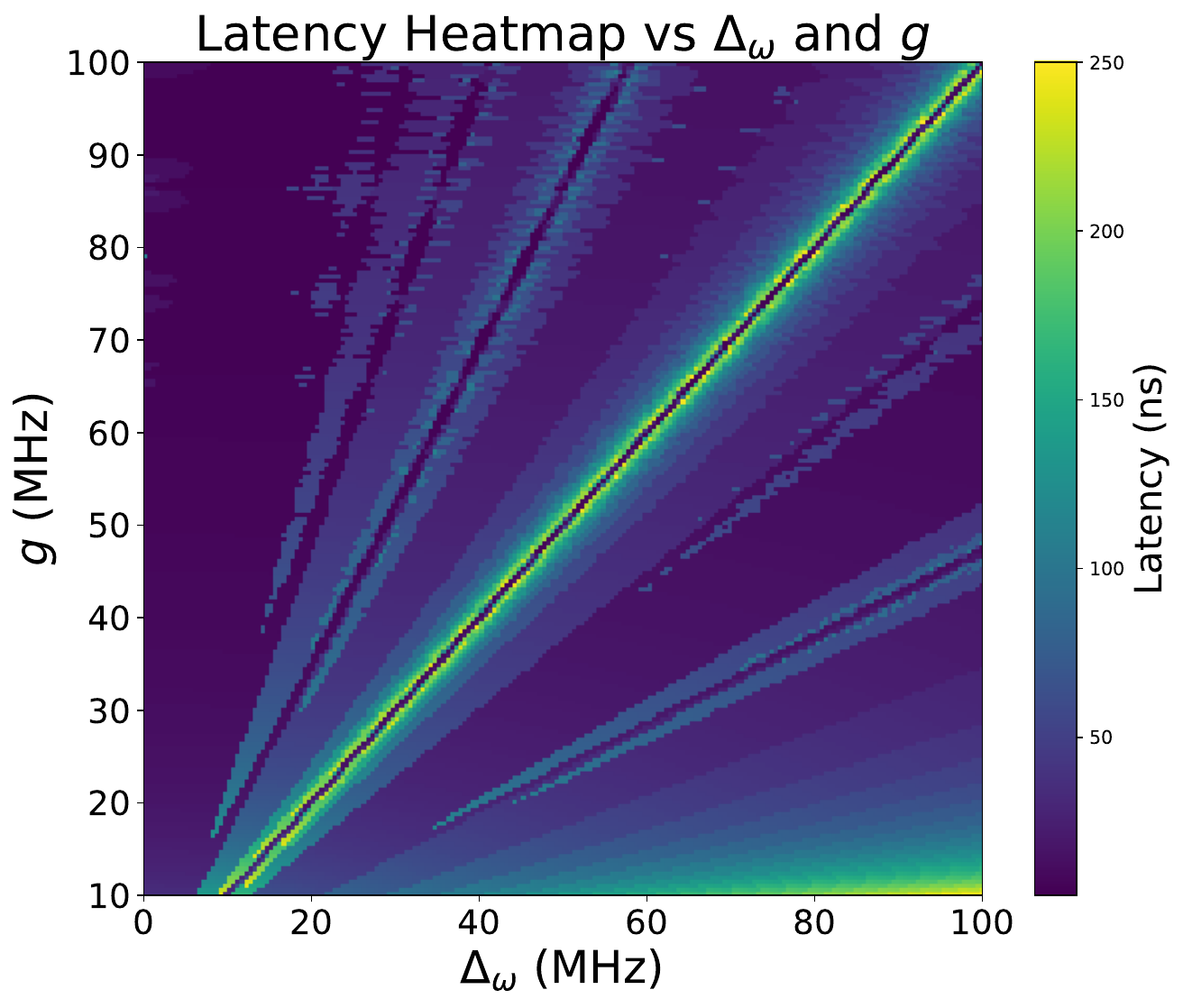}
        \caption{Latency for analytical equations.}
    \end{subfigure}
    \hfill
    \begin{subfigure}[t]{0.23\linewidth}
        \centering
        \includegraphics[width=\linewidth]{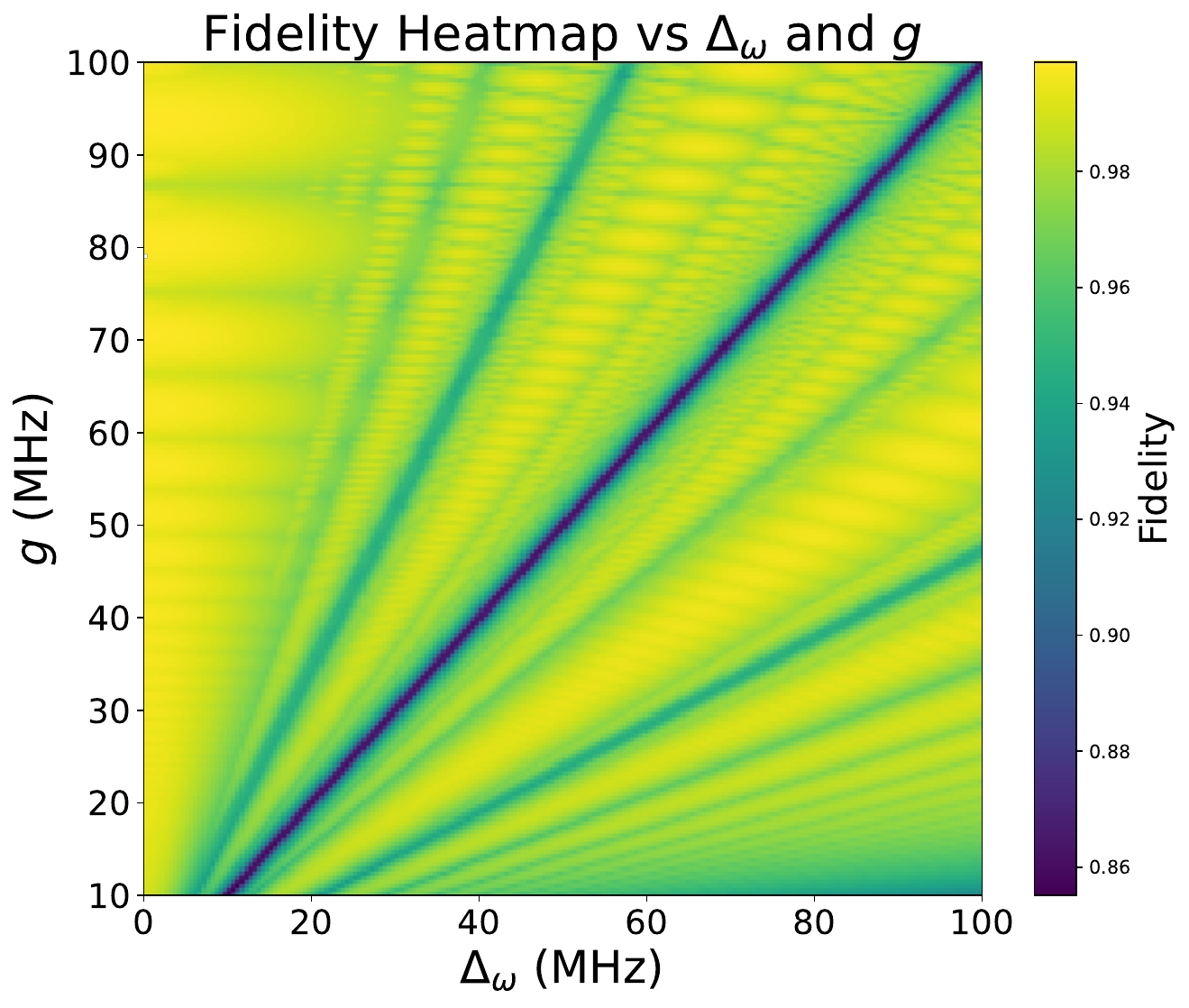}
        \caption{Fidelity for analytical equations.}
    \end{subfigure}
    \hfill
    \begin{subfigure}[t]{0.23\linewidth}
        \centering
        \includegraphics[width=\linewidth]{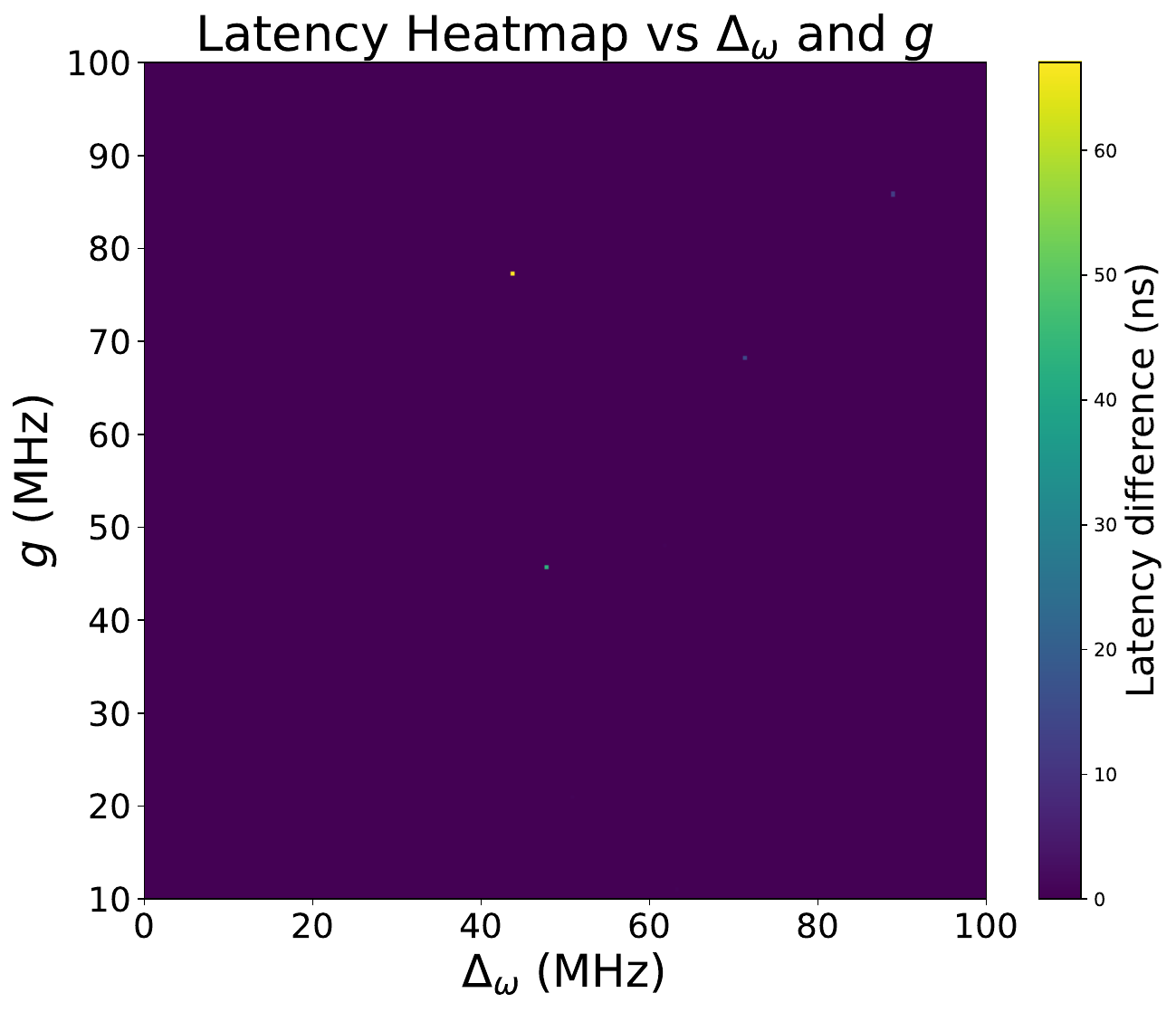}
        \caption{Difference between analytical and QuTiP latency.}
    \end{subfigure}
    \hfill
    \begin{subfigure}[t]{0.23\linewidth}
        \centering
        \includegraphics[width=\linewidth]{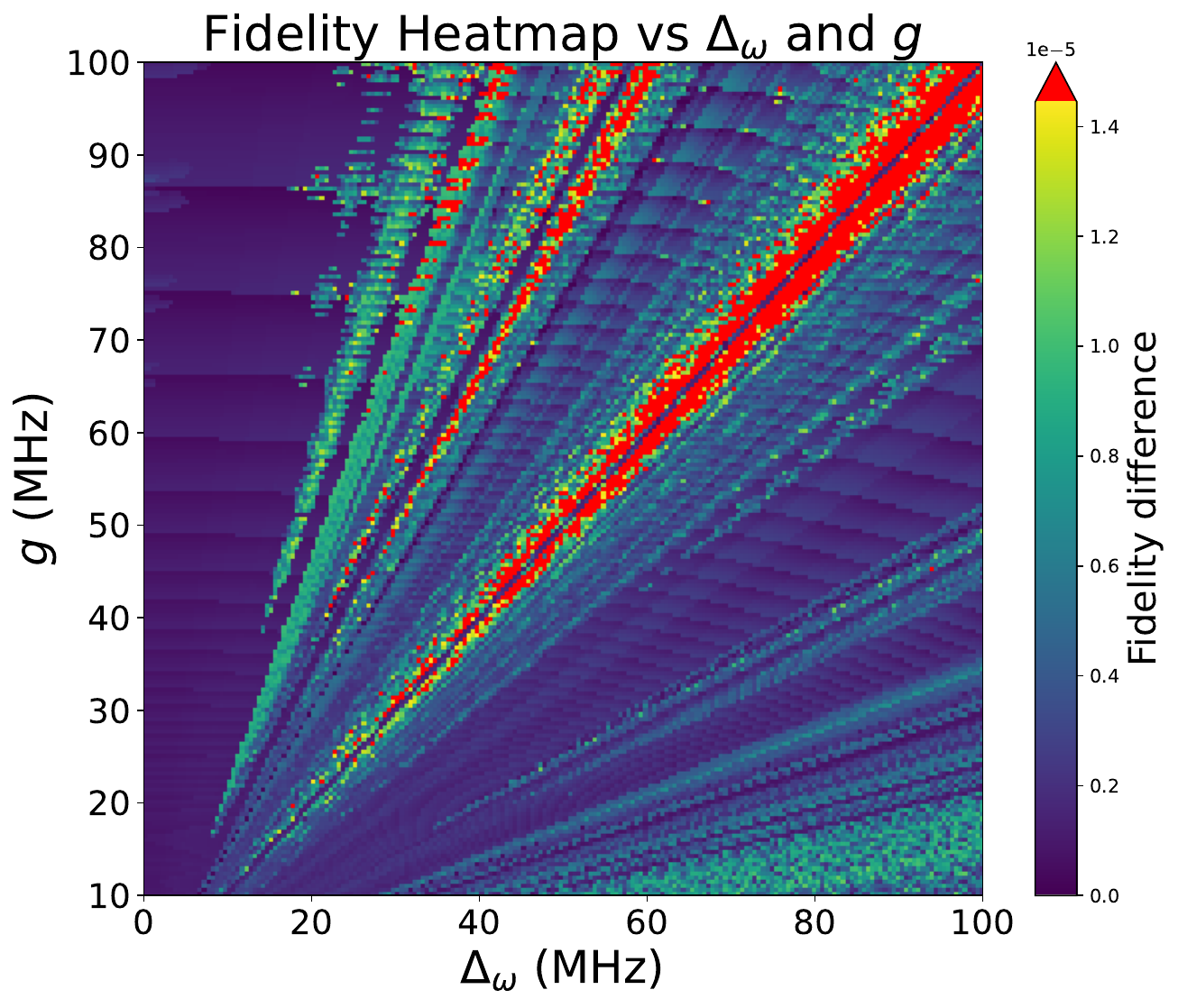}
        \caption{Difference between analytical and QuTiP fidelity, saturated at the 95th percentile.}
    \end{subfigure}
    \caption{Comparison between heatmaps using the analytical and simulation results for $\kappa = \gamma = 0.1$ MHz.}
    \label{fig:Fidelity Heatmap}
\end{figure*}
In Figure~\ref{fig:Fidelity Heatmap} we can observe a pattern in the appearance of low fidelity regions. Understanding such pattern is fundamental because one needs to avoid such regions in order to execute a state transfer efficiently. Notice that for small losses (up to $\sim$ 1 GHz), $\phi t \ll1$, so we can expand the equations \eqref{Lossy_Prob} to the first order of $\phi t$:
\begin{equation*}
\begin{aligned}
    P_{\text{WG}} (t) ={}& \frac{e^{-\frac{\kappa + \gamma}{2}t}}{2} \frac{\Omega^2}{\theta^2 + \phi^2} \left (\sin^2(\theta t) + 2\frac{\theta \phi}{\theta^2 + \phi^2} \sin(2 \delta t)\right ) \\
    &+\mathcal{O}(\phi t)^2
\end{aligned}
\end{equation*} 
\begin{figure*}[t]
    \centering
    \begin{subfigure}{0.35\linewidth}
        \centering
        \includegraphics[width=\linewidth]{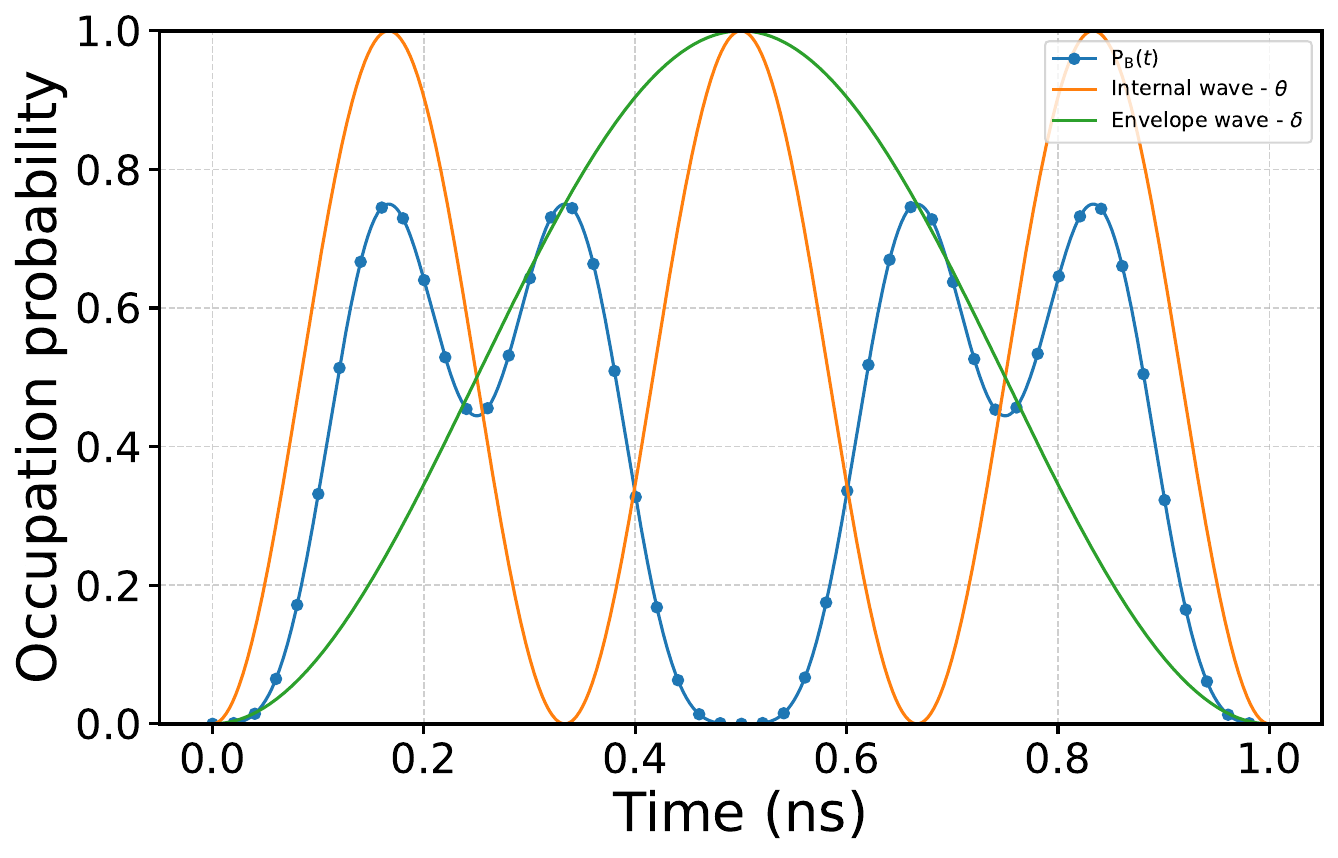}
        \caption{}
    \end{subfigure}
    \begin{subfigure}{0.35\linewidth}
        \centering
        \includegraphics[width=\linewidth]{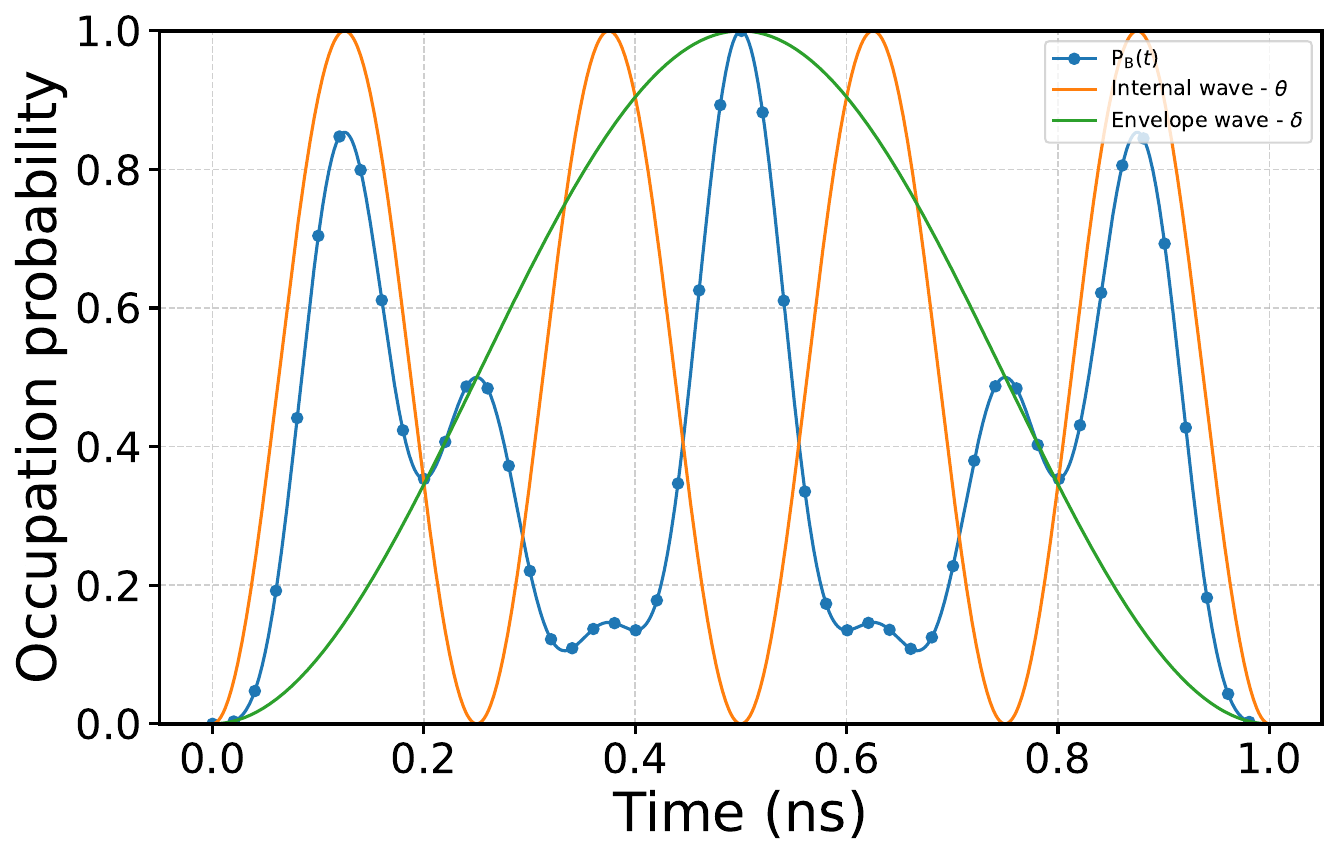}
        \caption{}
    \end{subfigure}
    \begin{subfigure}{0.28\linewidth}
        \centering
        \includegraphics[width=\linewidth]{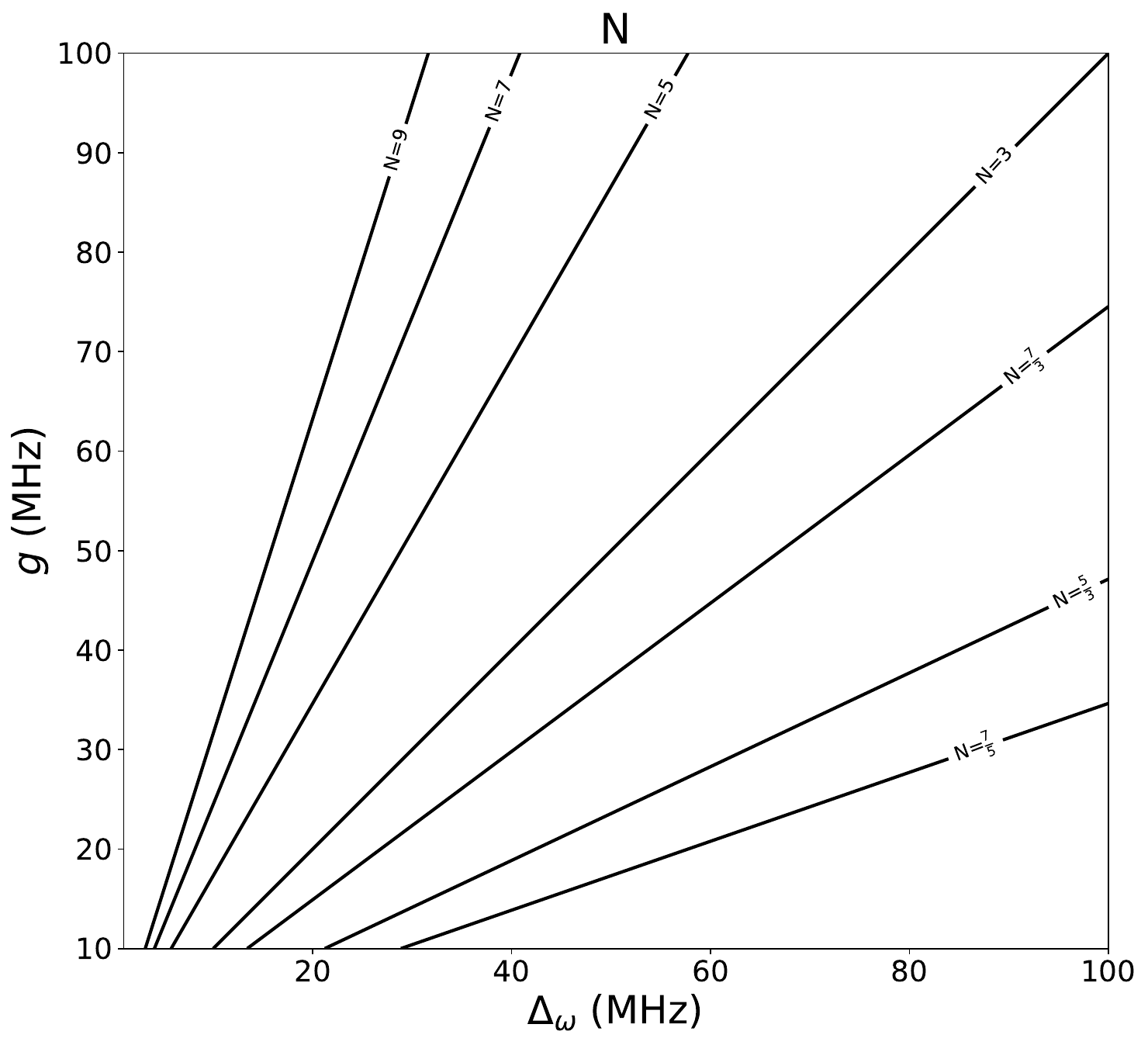}
        \caption{}
        \label{fig:N_map}
    \end{subfigure}
    \caption{Study of the types of $N$ combinations. If $a$ and $b$ are both odd, fidelity will always be low. If $N$ is even, fidelity will be high with low latency. a) Occupation probability for $N = 3$ for a lossless system. b) Occupation probability for $N = 4$ for a lossless system. c) Contour map of some relevant values of N for low fidelity regions due to the periodicity of the envelope and the internal waves, for $\kappa = 0.1$ MHz and $\gamma = 0.1$ MHz.}
    \label{N comparison}
\end{figure*}

Where $P_\text{WG} (t)$ is the probability of finding the Fock state in the Waveguide at time $t$. So, in a lossy system, the oscillations of the waveguide are $P_{\text{WG}} \sim \sin^2(\theta t) $, just like the lossless system. For the fidelity of the second qubit, we are interested in the instances where the waveguide is empty, so we are going to consider $\sin^2(\theta t) = 0$:
\begin{equation*}
\begin{aligned}
    P_B (t)|_{sin(\theta t) =0} ={} &\frac{1}{4} \Bigl(
    e^{-\gamma t}
    - 2 e^{-\frac{\kappa + 3\gamma}{4}t}
    [\,(1- a \phi t)\cos(\delta t)\\
    & - b \phi t\sin(\delta t)\,] \cos(\theta t)\\
    &+ e^{-\frac{\kappa+\gamma}{2}t}(1-2a\phi t)
    \Bigr)  + \mathcal{O}(\phi t)^2
    \label{first approximation}
\end{aligned}
\end{equation*}
And, to zero order,
\begin{equation*}
\begin{aligned}
        P_B (t)|_{sin(\theta t) =0} ={} &\frac{1}{4} \Bigl(
    e^{-\gamma t}
    - 2 \cos(\theta t)\cos(\delta t) e^{-\frac{\kappa + 3\gamma}{4}t}\\
    &+ e^{-\frac{\kappa+\gamma}{2}t}
    \Bigr)  + \mathcal{O}(\phi t)
\end{aligned}
\end{equation*}
While in a lossless system, we get
$$
P_B (t)|_{sin(\theta t) =0} =\frac{1}{2} \left(1- \cos(\theta t) \cos(\delta t)
\right)
$$
Therefore, the oscillations of the qubits behave like the product of two cosine waves, $P_B(t) \sim -\cos(\theta t) \cos(\delta t)$. Notice that, since $\theta > \delta$, the $\cos(\delta t)$ wave will behave like an envelope for the $\cos(\theta t)$ wave. So fidelity will depend on how these frequencies behave with each other. If we want to find cases with maximum $P_B (t)$, we must take $t$ where $\cos(\theta t) \cos(\delta t)$ is minimum. We define $N$ as the relation between the period of the envelope and the internal waves, $N = \frac{\theta}{\delta}$.
If $N$ is even, the occupation probability reaches its maximum at the maximum of the envelope, Fig.~\ref{N comparison}. But if $N = \frac{a}{b}$ with $a$ and $b$ both odd, the occupation probability can never reach 1 (i.e. $\cos(\theta t) \cos(\delta t) > 0 \ \forall t$), especially for small values of $a$ and $b$. In these regions we will always have low fidelity.

In Fig~\ref{N comparison}a, $\theta$ oscillations are three times faster than $\delta$ oscillations. As a result, the occupation probability $P_B$ is bounded by a maximum value of $3/4$. This value of $N$ corresponds to the low-fidelity main diagonal observed in Fig.~\ref{fig:Fidelity Heatmap}, where the effect is most noticeable. In contrast, Fig~\ref{N comparison}b shows the case where the $\theta$ oscillations are four times faster than $\delta$ oscillations. Here, the wave reaches its maximum exactly at the same time as $\delta$, resulting in a high fidelity region. This value of $N$ corresponds to the high-fidelity, low-latency region located at the left of the main diagonal in Fig.~\ref{fig:Fidelity Heatmap}.


Fig.~\ref{fig:N_map} shows some of the most relevant values of $N$ for low fidelity in a map of coupling strength ($g$) vs frequency difference i.e. detuning ($\Delta_{\omega}$). Around these regions we can always find better fidelity with a lot more latency, because the waves' periodicity don't match exactly. As a matter of fact, the time for optimal transmission in the product of the waves diverges around $N = \frac{a}{b}$ with $a$ and $b$ both odd, with $t \rightarrow \infty$ at the $N$. For this reason we can expect to find low-latency lines (with low fidelity) that match our $N$ plot, surrounded by high-latency regions on both sides (with higher but still low fidelity). This description perfectly matches the simulation obtained in Fig.~\ref{fig:Fidelity Heatmap}. 
\begin{figure*}[t]
    \centering
    \begin{subfigure}{0.23\linewidth}
        \centering
        \includegraphics[width=\linewidth]{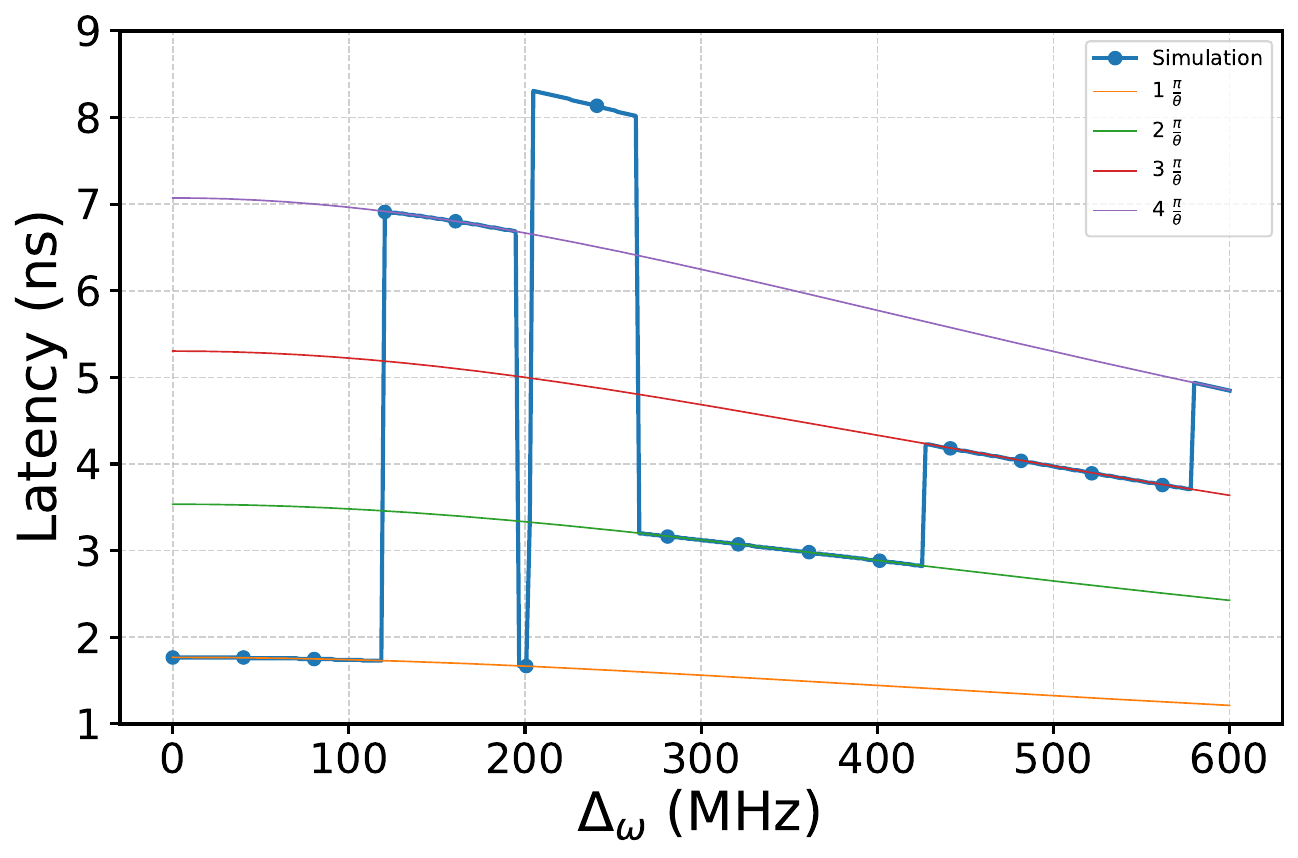}
        \caption{}
    \end{subfigure}
    \begin{subfigure}{0.23\linewidth}
        \centering
        \includegraphics[width=\linewidth]{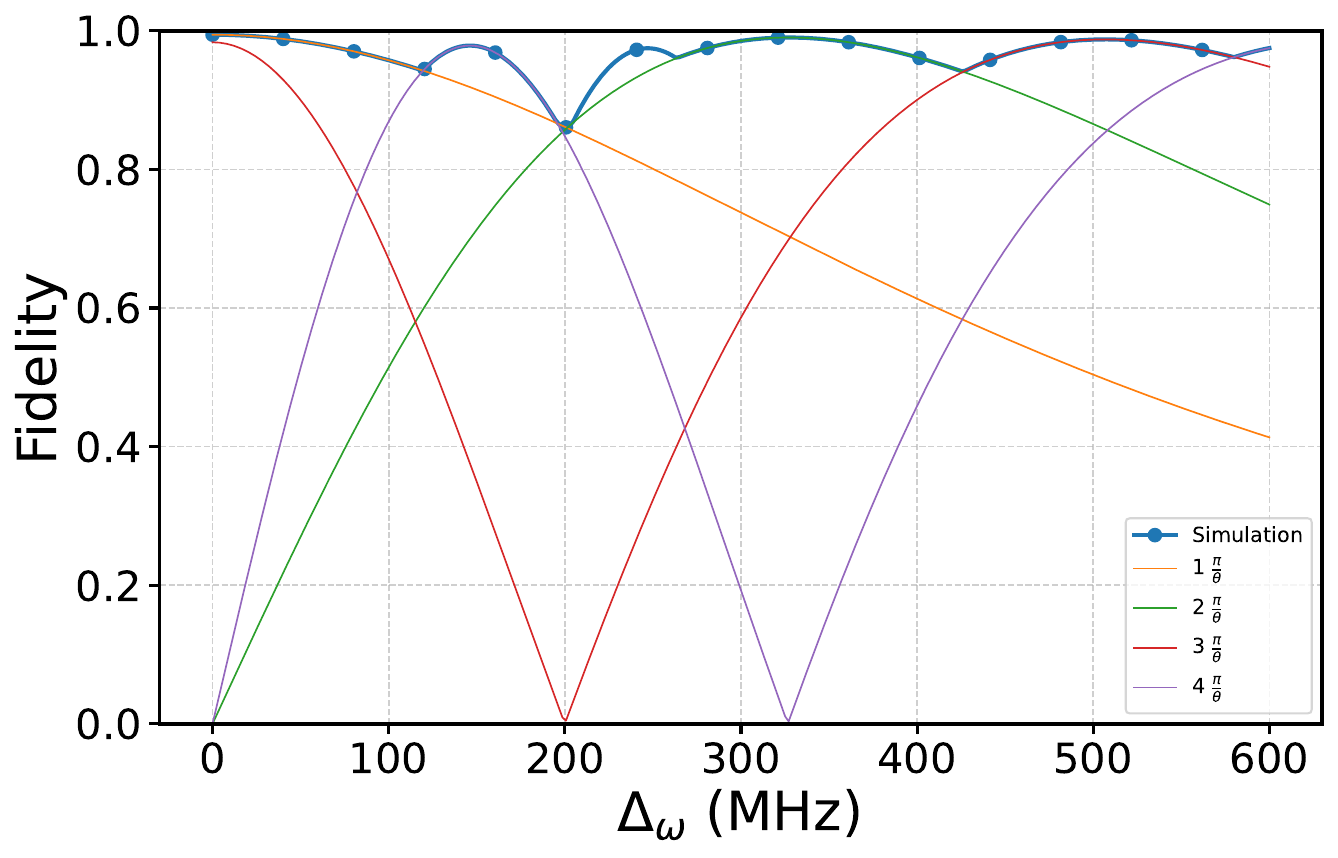}
        \caption{}
    \end{subfigure}
    \centering
    \begin{subfigure}{0.23\linewidth}
        \centering
        \includegraphics[width=\linewidth]{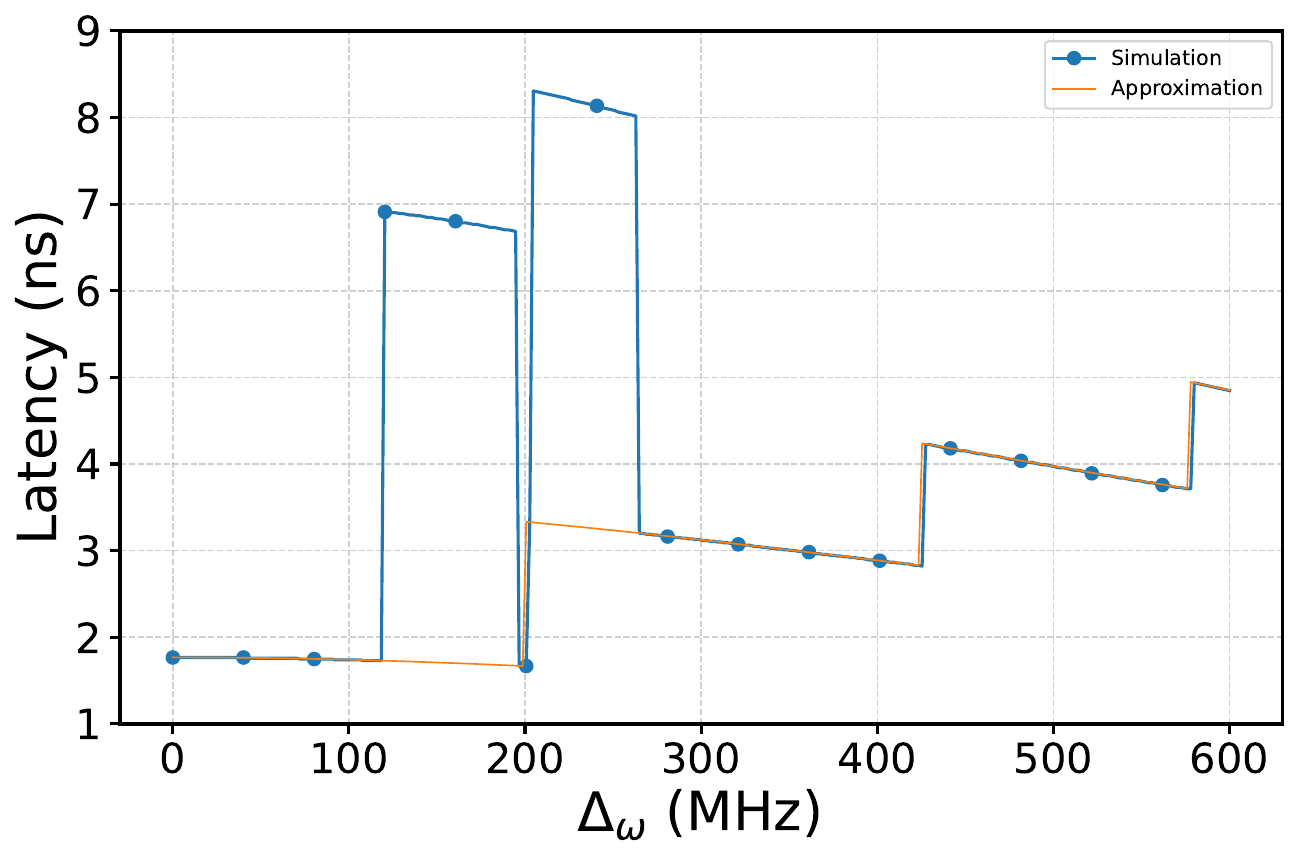}
        \caption{}
    \end{subfigure}
    \begin{subfigure}{0.23\linewidth}
        \centering
        \includegraphics[width=\linewidth]{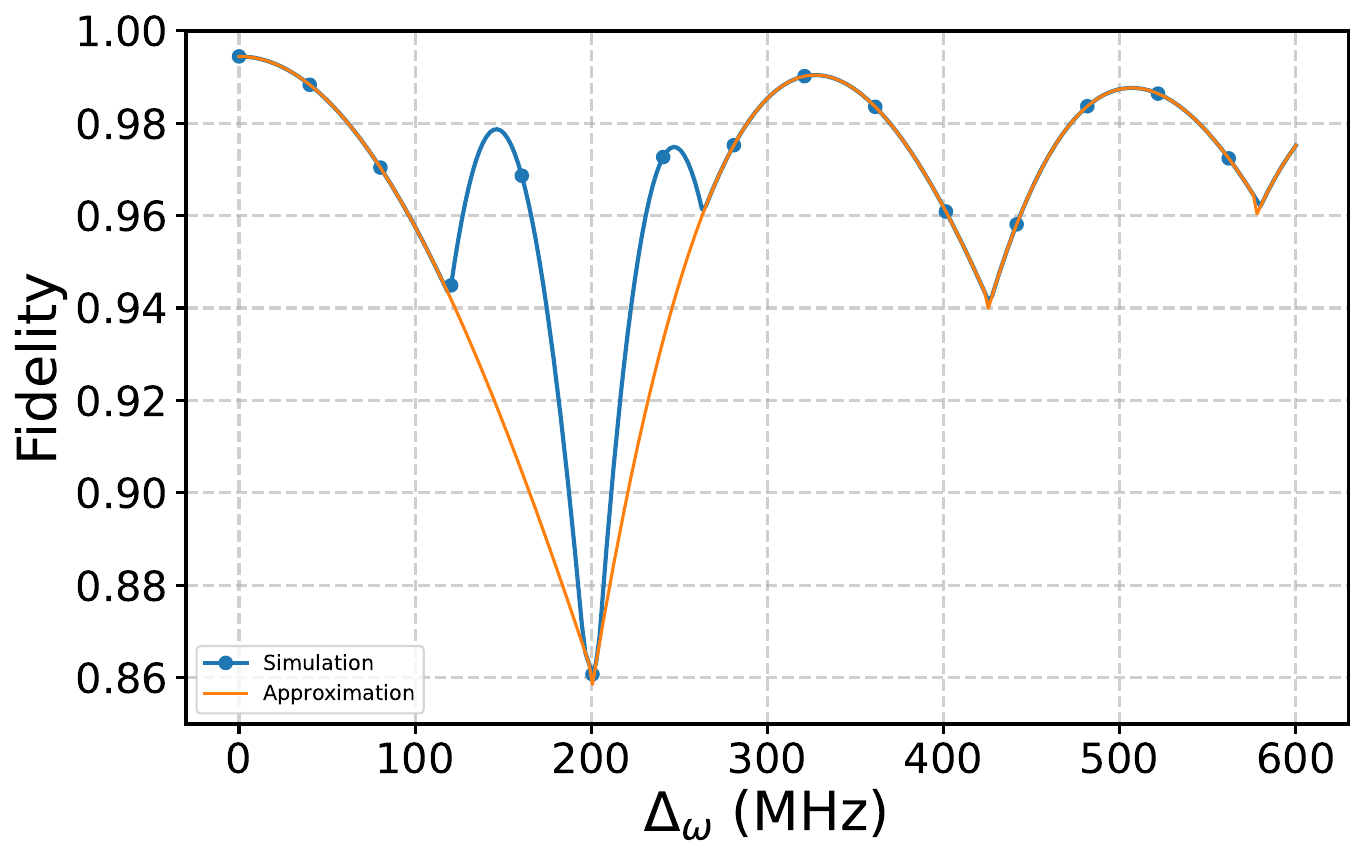}
        \caption{}
    \end{subfigure}
    \caption{Comparison of latency and fidelity for $g = 200$ MHz, $\kappa = 1$ MHz, and $\gamma = 1$ MHz.
a) Latency comparison between the simulation and the results at multiples of $\pi/\theta$; b) corresponding fidelity comparison. c) latency comparison between our model and the simulation; d) corresponding fidelity comparison.}
    \label{Model 1}
\end{figure*}

\begin{figure*}[t]
    \centering
    \begin{subfigure}{0.23\linewidth}
        \centering
        \includegraphics[width=\linewidth]{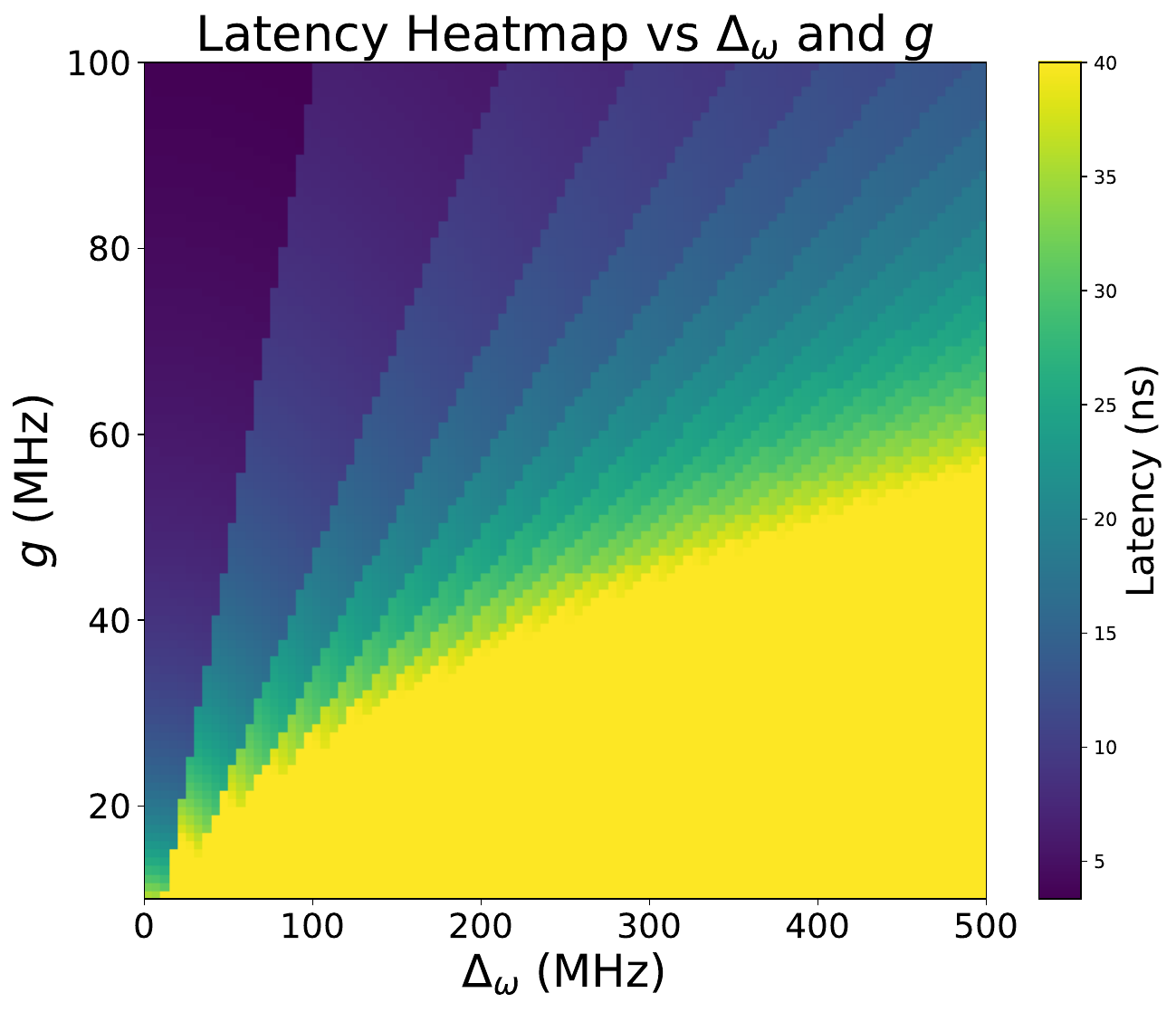}
        \caption{}
        \label{Model 2 a}
    \end{subfigure}
    \begin{subfigure}{0.23\linewidth}
        \centering
        \includegraphics[width=\linewidth]{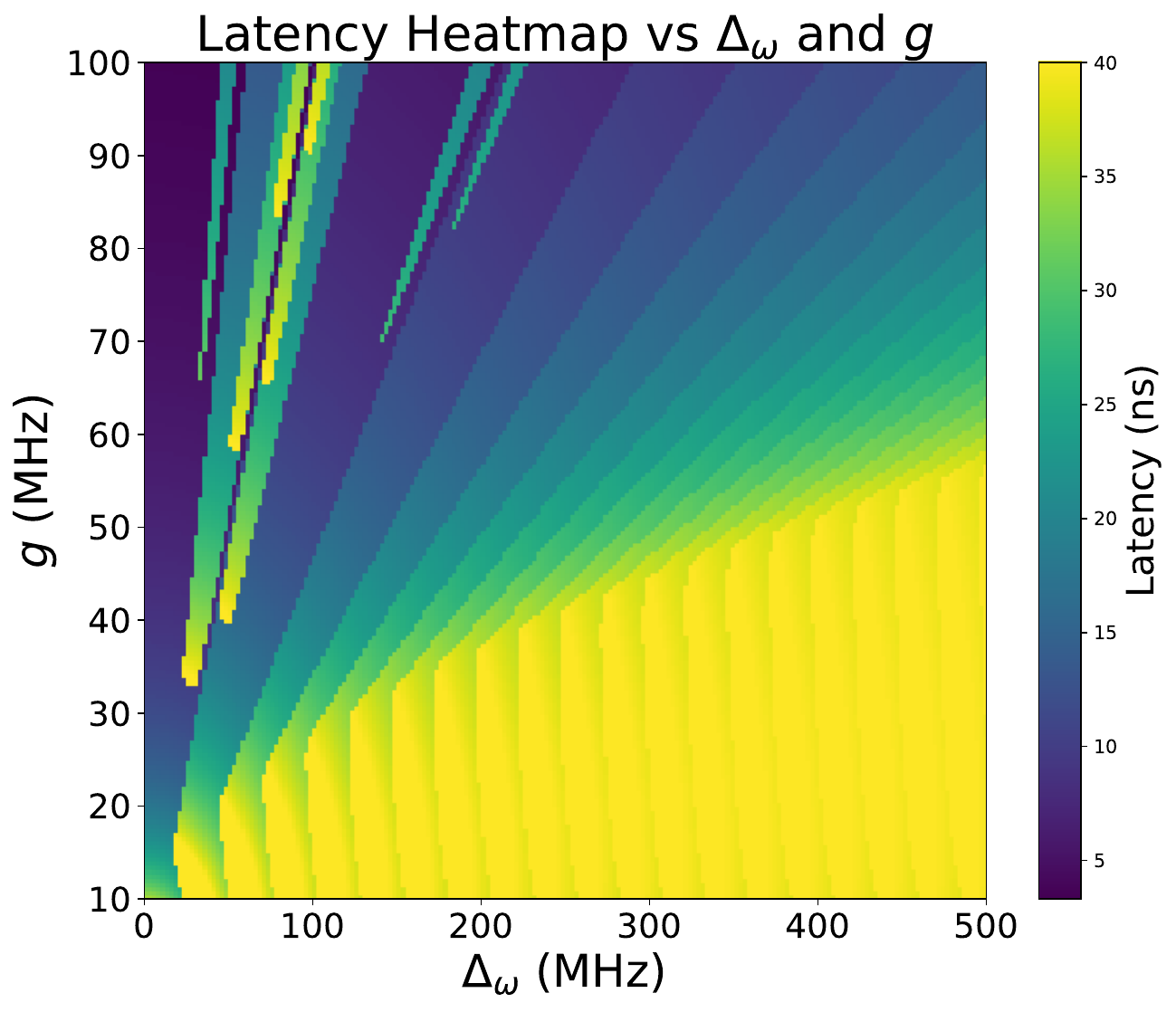}
        \caption{}
        \label{Model 2 b}
    \end{subfigure}
    \begin{subfigure}{0.23\linewidth}
        \centering
        \includegraphics[width=\linewidth]{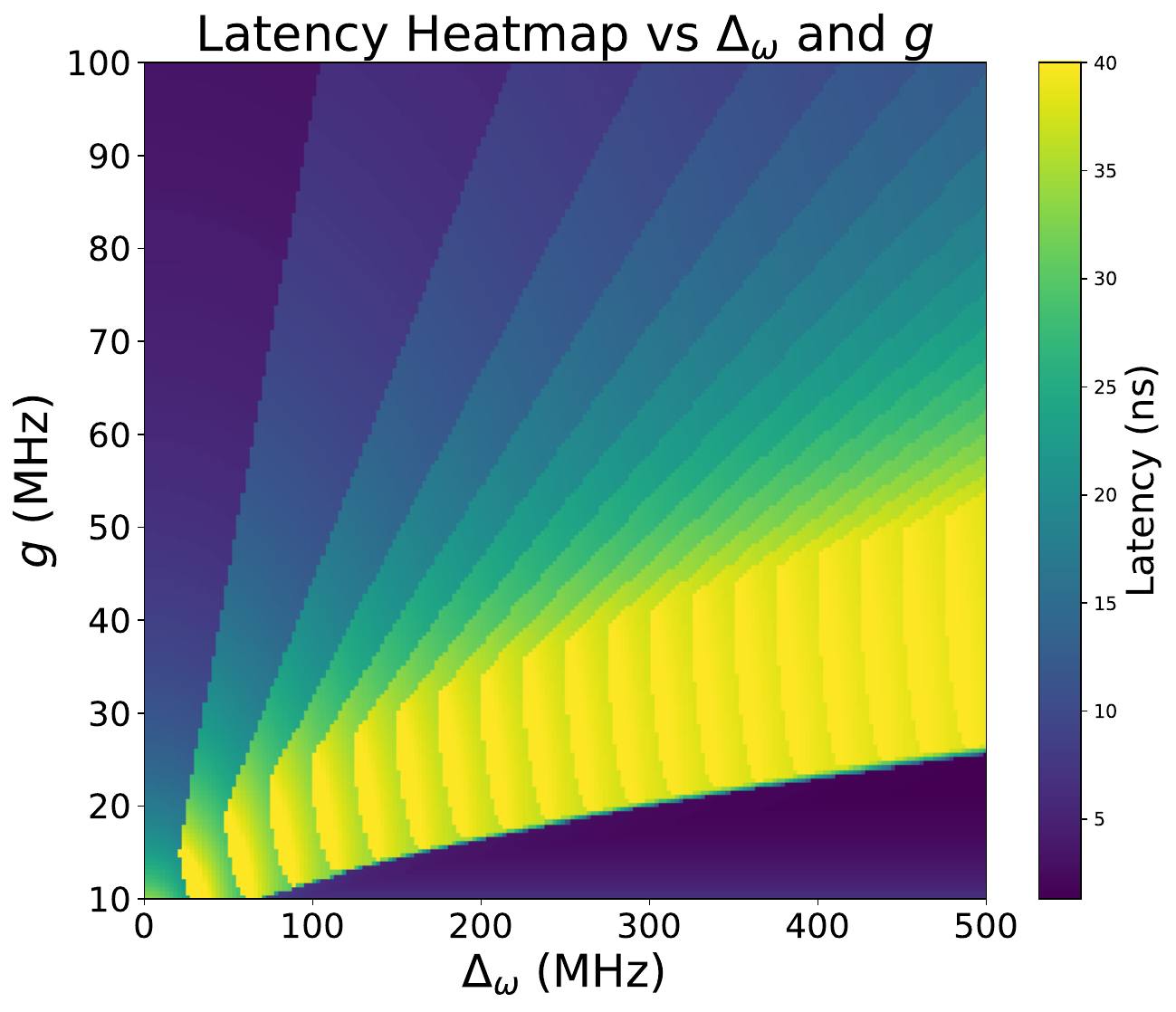}
        \caption{}
        \label{Model 2 c}
    \end{subfigure}
    \caption{Comparison of the latency heatmaps obtained with the analytical model of latency and the full QuTiP simulation, prioritizing short latencies over high fidelities. The cavity and qubit loss rates are set to $\kappa = \gamma = 0.1$~MHz. a) Analytical Model. b) QuTip simulation ($\eta = 1 \times 10^6$). c) QuTip simulation ($\eta = 8 \times 10^6$).}
    \label{Model 2}
\end{figure*}
\subsection{Latency and fidelity predictor model}
 Even with the analytical solution, a time sweep is still required to determine the optimal latency and fidelity. To address this, we have developed a model that allows the latency to be estimated using a simple analytical expression, thereby avoiding the need for a full-time sweep. For small losses, the frequency of the occupation probabilities behaves as the product of an envelope and an internal wave, so latency will be an integer multiple of the half-periods of the internal waves:
\begin{equation*}
    \tau = \mathcal{C} \ \frac{\pi}{\theta} \qquad \mathcal{C} \in \mathbb{N}
\end{equation*}
In Fig. \ref{Model 1}, we can see that latency is a multiple of $\frac{\pi}{\theta}$ as expected. Taking these considerations into account, a suitable model for determining the latency can be obtained by noting that a maximum occurs whenever the two waves (the envelope and the main oscillation) are in phase. Whether both are at a maximum or a minimum, their product attains a maximum in either case. Thus, we impose the condition that their arguments must either be equal or differ by $2\pi n$.
\begin{gather}
\theta t = \delta t + \pi + 2\pi n, \qquad
t = \frac{\pi}{\theta - \delta} \nonumber\\
\mathcal{C} = \frac{t}{\pi/\theta} =\frac{\theta}{\theta-\delta}, \qquad
\tau = \text{round}\left(\frac{\theta}{\theta-\delta}\right)\frac{\pi}{\theta}
\label{eq:latency prediction}
\end{gather}
Where we have rounded $\frac{\theta}{\theta - \delta}$ because $\mathcal{C}$ must be a natural number.


To check the computational cost and validity of our analytical latency model, Eq.~\eqref{eq:latency prediction}, we proceed in a similar manner as before. But in this case, the detuning $\Delta_{\omega}$ spans from $0$ to $500$~MHz in the heatmap. As shown in Fig.~\ref{Model 2}, the analytical model accurately reproduces the QuTiP results for small loss rates, up to $\gamma, \kappa \sim 1$~MHz, provided that the latency--fidelity trade-off parameter (explained in Appendix~\ref{sec:efficiency}) is chosen to be around $\eta \approx 10^6$. The benchmarking protocol is the same as before, but now time sweeps span from $0$ to $40$~ns for the QuTip simulation. Under these conditions, the total execution time amounts to \(T_{\mathrm{QuTiP}} = 1200 \pm 70\)~s for the QuTiP simulations, while the analytical model requires only \(T_{\mathrm{model}} = 0.974 \pm 0.004\)~s.
We can notice three main differences between the analytical model and the QuTiP simulations shown in Fig.~\ref{Model 2}:
\begin{enumerate}
    \item The appearance of latency oscillations in the lower part of the simulation heatmaps. For small coupling strengths, the optimal time for latency diverges. However, since the simulations are restricted to a finite time window (of $40\,\mathrm{ns}$, in our case), the obtained values remain within this range in the simulation.
    
    \item The appearance of higher latency regions in Fig.~\ref{Model 2 b}, in the upper left part of the heatmap. These regions are the 2D equivalence to the protuberances shown in Fig.~\ref{Model 1}.

    \item The appearance of a region where the latency is effectively zero in Fig.~\ref{Model 2 c}, in the lower right part of the heatmap. Due to the latency divergence for small $g$ and the chosen $\eta$ parameter, the \textit{efficiency function} $J(\eta, t)$ in Eq.~\eqref{eq:efficency} decreases with time in this region. So the optimal transmission for this region is considered to be with latency = fidelity = 0. 
\end{enumerate}
These differences arise from the different criteria used to define an \textit{optimal transmission}. In Fig.~\ref{Model 2 b}, higher fidelity is prioritized over low latency, whereas in Fig.~\ref{Model 2 c}, low latency is favored over higher fidelity, to the extent that transmission may not occur at all. The proposed model provides a general understanding of how the fidelity behaves for different combinations of parameters. This demonstrates that, for a given coupling strength, detuning increases the latency in the regime of small and similar losses. Hence, from this discussion, one can find an optimal region for quantum state-transfer in a waveguide model.

\section{Conclusions}
\label{Conclusion}
In this work, we derived analytical expressions for quantum state transfer through a waveguide, showing that the oscillation frequency $\theta$ is governed by detuning, coupling strength, and decay rates rather than the bare qubit or waveguide frequencies. In the small-loss regime, regions where $\theta/\delta$ takes fractional values with odd denominators should be avoided, and the latency grows with detuning following $\theta/(\theta - \delta)$. On the other hand, under strong waveguide losses, detuning improves fidelity. Finally, this model shows that waveguide losses drive it toward $1/2$ at long times and qubit losses suppress it entirely. The analytical model runs consistently faster than QuTiP simulations, making it suitable for rapid parameter exploration, and the approach extends naturally to any finite number of waveguide modes. 

\section{Acknowledgment}
This work was supported in part by the EC through HORIZON-EIC-2022-PATHFINDEROPEN-01-101099697 (QUADRATURE) and HORIZON-ERC-2021-101042080 (WINC). CGA acknowledges support from the Ministry for Digital Transformation and of Civil Service of the Spanish Government through the QUANTUM ENIA project call - Quantum Spain project, and by the European Union through the Recovery, Transformation and Resilience Plan - NextGenerationEU within the framework of the Digital Spain 2026 Agenda. Also, from the Spanish Ministry of Science, Innovation and Universities and European ERDF under grant PID2024-158682OB-C31. EA acknowledges support from Generalitat de Catalunya, ICREA Academia Award 2024. RSS acknowledges support from MSCA-COFUND Ramon Llull AIRA (HORIZON-MSCA-2022-cofund-01, Number: 101126667).

\appendix
\section{Annex}

\subsection{Deriving the equations for the lossless model}
\label{Lossless_equations}

To diagonalize the Hamiltonian of the a single-state excitation basis \eqref{eq:Lossless_Hamiltonian}, we must find the roots of the characteristic polynomial
\begin{equation}
(\omega_q - \lambda)^2 (\omega_{WG}- \lambda) - 2g^2 (\omega_q - \lambda) = 0
\end{equation}
where we have considered that both qubits share the same frequency $\omega_q$ and that the waveguide mode is detuned $\Delta_{\omega}$ by, such that $\omega_{WG} = \omega_q + \Delta_{\omega}$. Solving these equations, we find the following eigenvalues,
\[
\lambda_0 = \omega_q, \qquad
\lambda_+ = \omega_q + \delta + \theta, \qquad
\lambda_- = \omega_q + \delta - \theta
\]
Where we have defined $ \delta = \Delta_{\omega} / 2$ , $ \Omega = \sqrt{2} g $ and $ \theta = \sqrt{\Omega^2 + \delta^2}$. The corresponding eigenvectors are
\[
\left\{
\begin{aligned}
|\lambda_0\rangle &= \frac{|1,0,0\rangle - |0,0,1\rangle}{\sqrt{2}} \\
|\lambda_{\pm}\rangle &= 
\mathcal{N}_{\pm} \left(
|1,0,0\rangle 
+ \frac{\sqrt{2} (\delta \pm \theta)}{\Omega} |0,1,0\rangle
+ |0,0,1\rangle
\right)
\end{aligned}
\right.
\]
With a normalization constant
\[
\mathcal{N}_{\pm} = \frac{1}{2} \sqrt{ \frac{\theta \mp \delta}{\theta} } 
\]
We can invert the relations to find the original states
\[
\left\{
\begin{aligned}
|1,0,0\rangle &= \frac{1}{\sqrt{2}} |\lambda_0\rangle 
+ \mathcal{N}_+ |\lambda_+ \rangle  
+ \mathcal{N}_- |\lambda_- \rangle \\
|0,1,0\rangle &= 
\mathcal{N}_+ \frac{\sqrt{2} (\delta + \theta)}{\Omega} |\lambda_+ \rangle  
+ \mathcal{N}_-  \frac{\sqrt{2} (\delta - \theta)}{\Omega} |\lambda_- \rangle \\
|0,0,1\rangle &= -\frac{1}{\sqrt{2}} |\lambda_0\rangle 
+ \mathcal{N}_+ |\lambda_+ \rangle  
+ \mathcal{N}_- |\lambda_- \rangle
\end{aligned}
\right.
\]
The system evolves from the state $|1,0,0\rangle$ with a unitary evolution:
\begin{equation*}
    \begin{aligned}
|\psi(t)\rangle =&
\frac{1}{\sqrt{2}} |\lambda_0\rangle e^{-i \omega t}
+ \mathcal{N}_+ |\lambda_+\rangle e^{-i(\omega + \delta + \theta)t} \\
&+ \mathcal{N}_- |\lambda_-\rangle e^{-i(\omega + \delta - \theta)t}
    \end{aligned}
\end{equation*}
As an example, we can solve the probability amplitude of the second qubit, $P_B = |\langle 0,0,1 | \psi(t) \rangle|^2$.
We compute:
\begin{equation*}
\begin{aligned}
    \langle 0,0,1 | \psi(t) \rangle =&
\frac{e^{-i\delta t}}{4\theta}
\left(
(\theta - \delta) e^{-i\theta t}
+ (\theta + \delta) e^{i\theta t}
\right)
- \frac{1}{2}\\
\implies& \frac{1}{2}(\cos{\delta t} - i \sin{\delta t})\left ( \cos{\theta t}  + i\frac{\delta}{ \theta} \sin{\theta t} \right )  
\end{aligned}
\end{equation*}
Since, $P_B =  \langle 0,0,1 | \psi(t) \rangle$, we get:
\begin{equation*}
\left\{
\begin{aligned}
P_A =& \frac{1}{4} \Bigl( 
1 + \cos^2(\theta t) 
+ 2 \cos(\theta t)\cos(\delta t) \\
& + \frac{2 \delta}{\theta} \sin(\theta t)\sin(\delta t)
+ \frac{\delta^2 \sin^2(\theta t)}{\theta^2} 
\Bigr) \\
P_B =& \frac{1}{4} \Bigl( 
1 + \cos^2(\theta t) 
- 2 \cos(\theta t)\cos(\delta t) \\
& - \frac{2 \delta}{\theta} \sin(\theta t)\sin(\delta t)
+ \frac{\delta^2 \sin^2(\theta t)}{\theta^2} 
\Bigr) \\
P_{WG} =& \frac{1}{2} \frac{\Omega^2}{\theta^2} 
\sin^2(\theta t)
\end{aligned}
\right.
\end{equation*}
Since we are interested in the fidelity, we can focus on the situation where the waveguide is empty,  
i.e. $\theta t = \pi n$, so our probabilities can be approximated as:
\[
\left\{
\begin{aligned}
P_A &= \frac{1}{2} (1 + \cos(\theta t)\cos(\delta t)) \\
P_B &= \frac{1}{2} (1 - \cos(\theta t)\cos(\delta t))
\end{aligned}
\right.
\]

\subsection{Deriving the equations for the lossy model}
\label{Appendix:Lossy}
For our system, the non-Hermitian Hamiltonian in the single-excitation basis  
\(\{|1,0,0\rangle, |0,1,0\rangle, |0,0,1\rangle\}\) is:

\[
H_{\text{eff}} = 
\begin{pmatrix}
\omega_q - i\frac{\gamma}{2} & g & 0 \\
g & \omega_{WG} - i\frac{\kappa}{2} & g \\
0 & g & \omega_q - i\frac{\gamma}{2}
\end{pmatrix}
\]
Its eigenvalues are:
\[
\left\{
\begin{aligned}
\lambda_0 &= \omega_q - i\frac{\gamma}{2} \\
\lambda_{\pm} &= \omega_q + \delta 
- i\frac{\gamma+\kappa}{4} 
\pm \theta'
\end{aligned}
\right.
\]

\[
\delta = \Delta_{\omega} / 2 \qquad 
\Omega = \sqrt{2} g 
\qquad 
\Gamma = \frac{\gamma - \kappa}{4}
\]
\[
\theta' = \sqrt{\Omega^2 + \delta^2 - \Gamma^2 + i\,2\delta\Gamma}
\]
Notice that with a non-Hermitian Hamiltonian, sometimes $|\lambda \rangle \neq \langle \lambda | ^\dagger$, since they can have different eigenvalues. From now on, we are going to differentiate right bras and kets from left bras and kets:
\[
H_{\text{eff}} |\lambda^R \rangle = \lambda |\lambda^R \rangle, \qquad
\langle \lambda^L |H_{\text{eff}} = \lambda \langle \lambda^L|, \qquad
\langle \lambda^L | \neq |\lambda^R \rangle^\dagger
\]
With the following right eigenvectors:
\begin{equation*}
    \begin{aligned}
        \left\{
\begin{aligned}
|\lambda_0^R\rangle ={} & 
\frac{|1,0,0\rangle - |0,0,1\rangle}{\sqrt{2}} \\
|\lambda_{\pm}^R\rangle ={} &  \mathcal{N_{\pm}^R}\Bigl ( 
|1,0,0\rangle 
+ \sqrt{2}\frac{\delta + i\Gamma \pm \theta'}{\Omega}\, |0,1,0\rangle\\ &
+ |0,0,1\rangle \Bigr )
\end{aligned}
\right.
    \end{aligned}
\end{equation*}
Where $N_{\pm}^R$ are the normalization constants of the $|\lambda_{\pm}^R\rangle $ vectors. To find the left dual eigenvectors, we have to solve the following equation:
\[
H_{\text{eff}}^\dagger |\lambda^L \rangle = \lambda^* |\lambda^L \rangle, \qquad
|\lambda^L \rangle = \langle\lambda^L |^\dagger
\]
With the following left dual eigenvectors:
\begin{equation*}
    \begin{aligned}
        \left\{
\begin{aligned}
\langle \lambda_0^L| = {}& 
\frac{\langle 1,0,0| - \langle 0,0,1|}{\sqrt{2}} \\[6pt]
\langle \lambda_{\pm}^L| = {} & 
\mathcal{N}_\pm^L\!\Bigl(
\langle 1,0,0|
+ \sqrt{2}\frac{\delta + i\Gamma \pm \theta'}{\Omega}\,\langle 0,1,0| \\&
+ \langle 0,0,1|
\Bigr)
\end{aligned}
\right.
    \end{aligned}
\end{equation*}
Imposing orthonormalization conditions, $\langle \lambda^L_m |\lambda^R_n \rangle = \delta_{m,n}$,
we find that the normalization constants satisfy:
\[
\mathcal{N}_{\pm}^L\,\mathcal{N}_{\pm}^R
= \frac{1}{4}\,\frac{\Omega^2}{\theta'(\theta' \pm \delta \pm i\Gamma)}
\]
Notice that for the limit $\kappa, \gamma \rightarrow 0 $ we recover the lossless equations. The system evolves from the state \(|1,0,0\rangle\):
\begin{equation*}
    \begin{aligned}
    |\psi(t)\rangle
={} & \sum_n \langle \lambda^L_n | \psi(0)\rangle\,
|\lambda^R_n\rangle\, e^{-i\lambda_n t}
    \\
        |\psi(t)\rangle
={} & \frac{1}{\sqrt{2}} |\lambda_0\rangle e^{-i\omega_q t-\frac{\gamma}{2}t}
+ \mathcal{N}_+^L |\lambda_+\rangle e^{-i(\omega_q + \delta+\theta')t}\, e^{-\frac{\kappa+\gamma}{4}t}\\
&+ \mathcal{N}_-^L |\lambda_-\rangle e^{-i(\omega_q + \delta-\theta')t}\, e^{-\frac{\kappa+\gamma}{4}t}
    \end{aligned}
\end{equation*}
Since $e^{-i\omega_q t}$ is a general phase, it won't add any physical meaning to the equations.
\begin{equation*}
    \begin{aligned}
        |\psi(t)\rangle
={} &\frac{1}{\sqrt{2}} |\lambda_0\rangle e^{-\frac{\gamma}{2}t}
+ \mathcal{N}_+^L |\lambda_+\rangle e^{-i(\delta+\theta')t}\, e^{-\frac{\kappa+\gamma}{4}t}\\
&+ \mathcal{N}_-^L |\lambda_-\rangle e^{-i(\delta-\theta')t}\, e^{-\frac{\kappa+\gamma}{4}t}
    \end{aligned}
\end{equation*}
Expanding on the single-excitation basis:
\[
\begin{aligned}
|\psi(t)\rangle&  ={}
\frac{1}{2} (|1,0,0\rangle - |0,0,1\rangle) e^{-\frac{\gamma}{2}t}+ \mathcal{N}_+^L\mathcal{N}_+^R
(|1,0,0\rangle \\ &+ \sqrt{2}\frac{\delta + i\Gamma + \theta'}{\Omega}\, |0,1,0\rangle + |0,0,1\rangle)
e^{-i(\delta+\theta')t} e^{-\frac{\kappa+\gamma}{4}t}\\ & +
\mathcal{N}_-^L\mathcal{N}_-^R
(|1,0,0\rangle + \sqrt{2}\frac{\delta + i\Gamma - \theta'}{\Omega}\, |0,1,0\rangle \\ &+ |0,0,1\rangle)
e^{-i(\delta-\theta')t} e^{-\frac{\kappa+\gamma}{4}t}
\end{aligned}
\]
From this point, the probability of finding each subsystem in its excited state can be obtained by projecting the evolved state onto the corresponding single-excitation basis vectors. Since our focus is on the fidelity and latency of the second qubit, we explicitly derive the expressions for qubit B; the same procedure applies analogously to qubit A and to the waveguide mode. \vspace{-0.3cm}
\begin{equation*}
    \begin{aligned}
        \langle 0,0,1|\psi(t)\rangle
= {}& -\frac{e^{-\frac{\gamma}{2}t}}{2}
+\Bigl ( \mathcal{N}_+^L\mathcal{N}_+^R e^{-i(\delta + \theta' )t}+ \\ & \mathcal{N}_-^L\mathcal{N}_-^R e^{-i(\delta - \theta' )t} \Bigr ) e^{-\frac{\kappa + \gamma}{4} t}\\[4mm]
\implies{} & -\frac{e^{-\frac{\gamma}{2}t}}{2}
+\frac{\Omega^2}{4 \theta'} \Bigl (\frac{e^{i\theta' t}+e^{-i\theta' t}}{\Omega^2}\theta' \\ & + (\delta + i \Gamma)\frac{e^{i\theta' t}-e^{-i\theta' t}}{\Omega^2} \Bigr ) e^{-i \delta t}  e^{-\frac{\kappa + \gamma}{4} t}
    \end{aligned}
\end{equation*}
\[
\begin{aligned}
\implies{}
& -\frac{e^{-\frac{\gamma}{2}t}}{2}
+ \frac{e^{-\frac{\gamma+\kappa}{4}t}}{2} e^{-i\delta t}
\Bigl[
\cos(\theta' t) + \\ &(a+ib)i\sin(\theta' t)
\Bigr]
\end{aligned}
\]
where \(a+ib = \frac{\delta + i\Gamma}{\theta'}\). Since \(\theta'\) is complex, write:
\[
\theta' = \theta + i\phi,
\qquad
\theta = \text{Re}(\theta'),\ \phi = \text{Im}(\theta')
\]
\[
\theta = \sqrt{\frac{\sqrt{(\Omega^2 + \delta^2 + \Gamma^2)^2 - 4 \Omega^2 \Gamma^2} + \Omega^2 + \delta^2 - \Gamma^2}{2}}
\]
\[
\phi = \frac{\delta \Gamma}{|\delta \Gamma|} \sqrt{\frac{\sqrt{(\Omega^2 + \delta^2 + \Gamma^2)^2 - 4 \Omega^2 \Gamma^2} - \Omega^2 - \delta^2 + \Gamma^2}{2}}
\]
\begin{equation*}
\begin{aligned}
&\langle 0,0,1|\psi(t)\rangle
= -\frac{e^{-\frac{\gamma}{2}t}}{2}
+ \frac{e^{-\frac{\gamma+\kappa}{4}t}}{2} e^{-i\delta t}
\Bigl[
\cos(\theta t) \cosh(\phi t)
\\ &- i \sin(\theta t) \sinh(\phi t) 
+ (a+ib)\bigl(
i\sin(\theta t)\cos(\phi t)
- \cos(\theta t) \sinh(\phi t)
\bigr)
\Bigr]
\end{aligned}
\end{equation*}
\[
\langle 0,0,1|\psi(t)\rangle
= -\frac{e^{-\frac{\gamma}{2}t}}{2}
+ \frac{e^{-\frac{\gamma+\kappa}{4}t}}{2} (\cos(\delta t) - i \sin(\delta t))
\left[
A + iB\right]
\]
With:
\[
\left\{
\begin{aligned}
A (t) ={}& \cos(\theta t)\,\cosh(\phi t) 
     - a\cos(\theta t)\,\sinh(\phi t) \\ & - b\sin(\theta t)\,\cosh(\phi t) \\[2mm]
B (t) ={}& -\sin(\theta t)\,\sinh(\phi t) 
     + a\sin(\theta t)\,\cosh(\phi t) \\ &
     - b\cos(\theta t)\,\sinh(\phi t)
\end{aligned}
\right.
\]
\[
\begin{aligned} 
\text{Im}{\langle 0,0,1|\psi(t)\rangle} = {} & \frac{e^{-\frac{\gamma + \kappa}{4} t}}{2} (B \cos(\delta t) - A \sin(\delta t) )  
\end{aligned}
\]

Finally, we get the probabilities for each subsystem by calculating $|\langle 0,0,1|\psi(t)\rangle|$:
\begin{equation*}\label{Loss equations}
\left\{
\begin{aligned} 
P_A(t) ={} & \frac{1}{4} \Bigl(
e^{-\gamma t}
+ 2 e^{-\frac{\kappa + 3\gamma}{4}t}
[A\cos(\delta t) + B\sin(\delta t)] \\ &
+ e^{-\frac{\kappa+\gamma}{2}t}(A^2 + B^2)
\Bigr)\\[2mm]
P_B(t) = {}  & \frac{1}{4} \Bigl(
e^{-\gamma t}
- 2 e^{-\frac{\kappa + 3\gamma}{4}t}
[A\cos(\delta t) + B\sin(\delta t)] \\ &
+ e^{-\frac{\kappa+\gamma}{2}t}(A^2 + B^2)
\Bigr)\\[2mm]
P_{\text{WG}}(t) ={} & \frac{e^{-\frac{\kappa + \gamma}{2}t}}{2} \frac{\Omega^2}{\theta^2 + \phi^2} \Bigl ( 
\sin^2(\theta t) \cosh^2(\phi t) \\ & +\cos^2(\theta t) \sinh^2(\phi t) [2 \sin^2(\delta t)-1] \\ & + 2\frac{\theta \phi}{\theta^2 + \phi^2} \sin(2\delta t)
\Bigr)
\end{aligned}
\right.
\end{equation*}
The factor \(A^2 + B^2\) can also be written as:
\[
\begin{aligned}
    &A^2 + B^2=  1 + (a^2 + b^2 +1) \sinh^2(\phi t) \\& + (a^2 + b^2 -1) \sin^2(\theta t)\sinh(2 \phi t) - b\sin(2 \theta t)
\end{aligned}
\]

\subsection{Proof of single excitation basis}
\label{single_excitation}
If the cavity has $N$ Fock states, the Hamiltonian can be written in the following basis:
\begin{equation*}
    \begin{aligned}
        &\{|1,0,0\rangle, |0,1,0\rangle, |0,0,1\rangle, \quad \\&|1,1,0\rangle, |0,2,0\rangle, |0,1,1\rangle , \\ &\lil  ..., |1,N-1,0\rangle, |0,N,0\rangle, |0,N-1,1\rangle\}
\end{aligned}
\end{equation*}
And the effective Hamiltonian takes the form
\[
H_\text{eff} =
\begin{pmatrix}
H_0 & 0  &  & 0 \\
0 & H_1  & ... & 0 \\
 & ... &    & ...\\
0 & 0 & ... & H_N \\
\end{pmatrix}
\]
With
\[
H_{\text{k}} = 
\begin{pmatrix}
\omega_q - i\frac{\gamma}{2} & \sqrt{k+1} \ g & 0 \\
\sqrt{k+1} \ g & (k+1)(\omega_{WG} - i\frac{\kappa}{2}) & \sqrt{k+1} \ g \\
0 & \sqrt{k+1} \ g & \omega_q - i\frac{\gamma}{2}
\end{pmatrix}
\]
This Hamiltonian is composed of N separate subsystems, where, if the initial state belongs to one subsystem, it won't reach another one. In our case, the system begins with the cavity empty, so we can work with just the $H_0$ subsystem. This means that of all the states of the electromagnetic mode of the waveguide, only two are relevant: the ground state $|0\rangle$ and the single excitation state $|1\rangle$.

\subsection{Quantum Jumps}
\label{Appendix:Jumps}
The probability of jumping (probability of emitting a photon between \(t\) and \(t + \delta t\)) at time \(t\) is:
\[
\delta p(t) = \delta t \, \langle \psi(t) |  L_k^{\dagger} L_k| \psi(t) \rangle.
\]
For our case:
\[
\delta p(t) = \delta t \cdot [\, \gamma P_A(t) + \gamma P_B(t) + \kappa\, P_{WG}(t)\, ].
\]
Even though $p(t) \ll 1$ and decreases over time, the probability that at least one quantum jump occurs within the interval [0,t] increases with time. So it isn't necessarily negligible. However, as discussed in \textit{Mølmer et al.} \cite{Molmer93}, our wave function will not change in quantum jumps, since the initial state of the waveguide, $|0\rangle$, is an eigenstate of the corresponding ladder operator. For this reason, we do not consider quantum jumps.

\subsection{Efficiency equation}
\label{sec:efficiency}
For optimal transmission, a perceptible goal is to achieve both low latency and high fidelity. However, in the regime of sufficiently small losses, once a time instance with good fidelity is identified, it is often possible to find a later instance with slightly improved fidelity. This trade-off is typically not advantageous, but ultimately depends on the design criteria and system requirements. For an optimal transmission, we want to maximize fidelity while minimizing latency at the same time. To do so, we have proposed a function $J(\eta,t)$ that we will call \textit{efficiency}, which has to be maximized to get optimal transmission.  
\begin{equation}
    \label{eq:efficency}
J(\eta, t) = P_B(t) - \eta t
\end{equation}
Where $\eta$ is a parameter that allows us to give more relevance to high fidelity (low $\eta$) or to low latency (high $\eta$). For a given $\eta$ we will take latency ($\tau$) as $t$ of maximum $J(\eta, t)$, and fidelity as $P_B (\tau)$. For example, in Fig.\ref{fig:Fidelity Heatmap} it's used $\eta = 10^4$, in Fig. \ref{probability comparison}, if we take $\eta = 10^7$, we will get latency to be $\tau = 1.904 $ ns, but for $\eta = 10^8$, we get $\tau = 0.315$ ns.
\FloatBarrier
\bibliographystyle{IEEEtran}  
\bibliography{references}

@misc{Castin,
      title={A Wave Function approach to dissipative processes}, 
      author={Yvan Castin and Jean Dalibard and Klaus Molmer},
      year={2008},
      eprint={0805.4002},
      archivePrefix={arXiv},
      primaryClass={quant-ph},
}

@article{Molmer93,
author = {Klaus M{\o}lmer and Yvan Castin and Jean Dalibard},
journal = {J. Opt. Soc. Am. B},
number = {3},
pages = {524--538},
publisher = {Optica Publishing Group},
title = {Monte Carlo wave-function method in quantum optics},
volume = {10},
month = {Mar},
year = {1993},
doi = {10.1364/JOSAB.10.000524},
}

@article{reiserer2015cavity,
  title={Cavity-based quantum networks with single atoms and optical photons},
  author={Reiserer, Andreas and Rempe, Gerhard},
  journal={Reviews of Modern Physics},
  volume={87},
  number={4},
  pages={1379--1418},
  year={2015},
  publisher={APS}
}

@article{manzano2020short,
  title={A short introduction to the Lindblad master equation},
  author={Manzano, Daniel},
  journal={Aip advances},
  volume={10},
  number={2},
  year={2020},
  publisher={AIP Publishing}
}

@article{Cirac_1997,
   title={Quantum State Transfer and Entanglement Distribution among Distant Nodes in a Quantum Network},
   volume={78},
   ISSN={1079-7114},
   number={16},
   journal={Physical Review Letters},
   publisher={American Physical Society (APS)},
   author={Cirac, J. I. and Zoller, P. and Kimble, H. J. and Mabuchi, H.},
   year={1997},
   month=apr, pages={3221–3224} }

@article{bravyi2022future,
  title={The future of quantum computing with superconducting qubits},
  author={Bravyi, Sergey and Dial, Oliver and Gambetta, Jay M and Gil, Dar{\'\i}o and Nazario, Zaira},
  journal={Journal of Applied Physics},
  volume={132},
  number={16},
  year={2022},
  publisher={AIP Publishing}
}

@article{field2023,
  author  = {M. Field and others},
  title   = {Modular superconducting qubit architecture 
             with a multi-chip tunable coupler},
  journal = {arXiv:2308.09240},
  year    = {2023}
}

@inproceedings{khan2025,
  author    = {J. Khan and S. Navarro Reyes and 
               S. Ben Rached and E. Alarc\'{o}n and 
               P. Haring Bol\'{i}var and 
               C. G. Almud\'{e}ver and S. Abadal},
  title     = {Waveguide {QED} analysis of 
               quantum-coherent links for modular 
               quantum computing},
  booktitle = {2025 IEEE Int. Symp. Circuits and 
               Systems (ISCAS)},
  year      = {2025}
}

@article{cosme2018,
  author  = {J. M. G. Cosme and others},
  title   = {Interplay between speed and fidelity in 
             off-resonant quantum-state transfer 
             protocols},
  journal = {Physical Review A},
  year    = {2018}
}

@inproceedings{rached2024benchmarking,
  title={Benchmarking emerging cavity-mediated quantum interconnect technologies for modular quantum computers},
  author={Rached, Sahar Ben and Reyes, Sergio Navarro and Khan, Junaid and Almud{\'e}ver, Carmen G and Alarc{\'o}n, Eduard and Abadal, Sergi},
  booktitle={2024 IEEE International Conference on Quantum Computing and Engineering (QCE)},
  volume={1},
  pages={1908--1913},
  year={2024},
  organization={IEEE}
}

@article{soh2021high,
  title={High-fidelity state transfer between leaky quantum memories},
  author={Soh, Daniel and Chatterjee, Eric and Eichenfield, Matt},
  journal={Physical Review Research},
  volume={3},
  number={3},
  pages={033027},
  year={2021},
  publisher={APS}
}

@article{rodrigo2023,
  author  = {S. Rodrigo and others},
  title   = {Characterizing the inter-core qubit 
             traffic in large-scale quantum modular 
             architectures},
  journal = {IEEE Access},
  year    = {2024}
}

@inproceedings{Azem2016JaynesCummingsM,
  title={Jaynes-Cummings Model},
  author={Adan Azem},
  year={2016},
}

@article{stute2012tunable,
  title={Tunable ion--photon entanglement in an optical cavity},
  author={Stute, Andreas and Casabone, Bernardo and Schindler, Philipp and Monz, Thomas and Schmidt, Piet O and Brandst{\"a}tter, Birgit and Northup, Tracy E and Blatt, Rainer},
  journal={Nature},
  volume={485},
  number={7399},
  pages={482--485},
  year={2012},
  publisher={Nature Publishing Group UK London}
}

@article{jnane2022multicore,
  title={Multicore quantum computing},
  author={Jnane, Hamza and Undseth, Brennan and Cai, Zhenyu and Benjamin, Simon C and Koczor, B{\'a}lint},
  journal={Physical review applied},
  volume={18},
  number={4},
  pages={044064},
  year={2022},
  publisher={APS}
}

@article{niu2024multi,
  title={Multi-qubit dynamical decoupling for enhanced crosstalk suppression},
  author={Niu, Siyuan and Todri-Sanial, Aida and Bronn, Nicholas T},
  journal={Quantum Science and Technology},
  volume={9},
  number={4},
  pages={045003},
  year={2024},
  publisher={IOP Publishing}
}

@article{Jozsa01121994,
author = {Richard Jozsa},
title = {Fidelity for Mixed Quantum States},
journal = {Journal of Modern Optics},
volume = {41},
number = {12},
pages = {2315--2323},
year = {1994},
publisher = {Taylor \& Francis},
doi = {10.1080/09500349414552171}}

@article{rached2025characterizing,
  title={Characterizing the inter-core qubit traffic in large-scale quantum modular architectures},
  author={Rached, Sahar Ben and Agudo, Isaac Lopez and Rodrigo, Santiago and Bandic, Medina and Garcia-Saez, Artur and Feld, Sebastian and Van Someren, Hans and Alarc{\'o}n, Eduard and Almud{\'e}ver, Carmen G and Abadal, Sergi},
  journal={IEEE access},
  year={2025},
  publisher={IEEE}
}

@article{ang2024arquin,
  title={Arquin: Architectures for multinode superconducting quantum computers},
  author={Ang, James and Carini, Gabriella and Chen, Yanzhu and Chuang, Isaac and Demarco, Michael and Economou, Sophia and Eickbusch, Alec and Faraon, Andrei and Fu, Kai-Mei and Girvin, Steven and others},
  journal={ACM Transactions on Quantum Computing},
  volume={5},
  number={3},
  pages={1--59},
  year={2024},
  publisher={ACM New York, NY}
}

@article{Dirac_1927,
    author = {Dirac, Paul Adrien Maurice},
    title = {The quantum theory of the emission and absorption of radiation},
    journal = {Proceedings of the Royal Society of London. Series A, Containing Papers of a Mathematical and Physical Character},
    volume = {114},
    number = {767},
    pages = {243-265},
    year = {1927},
    month = {03},
    issn = {0950-1207},
    doi = {10.1098/rspa.1927.0039},
}

@article{preskill2018quantum,
  title={Quantum computing in the NISQ era and beyond},
  author={Preskill, John},
  journal={Quantum},
  volume={2},
  pages={79},
  year={2018},
  publisher={Verein zur F{\"o}rderung des Open Access Publizierens in den Quantenwissenschaften}
}

@article{niu2023low,
  title={Low-loss interconnects for modular superconducting quantum processors},
  author={Niu, Jingjing and Zhang, Libo and Liu, Yang and Qiu, Jiawei and Huang, Wenhui and Huang, Jiaxiang and Jia, Hao and Liu, Jiawei and Tao, Ziyu and Wei, Weiwei and others},
  journal={Nature Electronics},
  volume={6},
  number={3},
  pages={235--241},
  year={2023},
  publisher={Nature Publishing Group UK London}
}

@article{Schrodinger,
  title = {An Undulatory Theory of the Mechanics of Atoms and Molecules},
  author = {Schr\"odinger, E.},
  journal = {Phys. Rev.},
  volume = {28},
  issue = {6},
  pages = {1049--1070},
  numpages = {0},
  year = {1926},
  month = {Dec},
  publisher = {American Physical Society},
  doi = {10.1103/PhysRev.28.1049}
}

@inproceedings{escofet2023interconnect,
  title={Interconnect fabrics for multi-core quantum processors: A context analysis},
  author={Escofet, Pau and Rached, Sahar Ben and Rodrigo, Santiago and Almudever, Carmen G and Alarc{\'o}n, Eduard and Abadal, Sergi},
  booktitle={Proceedings of the 16th International Workshop on Network on Chip Architectures},
  pages={34--39},
  year={2023}
}

@article{llewellyn2020chip,
  title={Chip-to-chip quantum teleportation and multi-photon entanglement in silicon},
  author={Llewellyn, Daniel and Ding, Yunhong and Faruque, Imad I and Paesani, Stefano and Bacco, Davide and Santagati, Raffaele and Qian, Yan-Jun and Li, Yan and Xiao, Yun-Feng and Huber, Marcus and others},
  journal={Nature Physics},
  volume={16},
  number={2},
  pages={148--153},
  year={2020},
  publisher={Nature Publishing Group UK London}
}

@article{zhong2021deterministic,
  title={Deterministic multi-qubit entanglement in a quantum network},
  author={Zhong, Youpeng and Chang, Hung-Shen and Bienfait, Audrey and Dumur, {\'E}tienne and Chou, Ming-Han and Conner, Christopher R and Grebel, Joel and Povey, Rhys G and Yan, Haoxiong and Schuster, David I and others},
  journal={Nature},
  volume={590},
  number={7847},
  pages={571--575},
  year={2021},
  publisher={Nature Publishing Group UK London}
}

@article{magnard2020microwave,
  title={Microwave quantum link between superconducting circuits housed in spatially separated cryogenic systems},
  author={Magnard, Paul and Storz, Simon and Kurpiers, Philipp and Sch{\"a}r, Josua and Marxer, Fabian and L{\"u}tolf, Janis and Besse, Jean-Claude and Gabureac, Mihai and Reuer, Kevin and Akin, Abdulkadir and others},
  journal={arXiv preprint arXiv:2008.01642},
  year={2020}
}

@article{blais2021circuit,
  title={Circuit quantum electrodynamics},
  author={Blais, Alexandre and Grimsmo, Arne L and Girvin, Steven M and Wallraff, Andreas},
  journal={Reviews of Modern Physics},
  volume={93},
  number={2},
  pages={025005},
  year={2021},
  publisher={APS}
}

@article{johansson2012qutip,
  title={QuTiP: An open-source Python framework for the dynamics of open quantum systems},
  author={Johansson, J Robert and Nation, Paul D and Nori, Franco},
  journal={Computer physics communications},
  volume={183},
  number={8},
  pages={1760--1772},
  year={2012},
  publisher={Elsevier}
}

@article{molmer1993monte,
  title={Monte Carlo wave-function method in quantum optics},
  author={M{\o}lmer, Klaus and Castin, Yvan and Dalibard, Jean},
  journal={Journal of the Optical Society of America B},
  volume={10},
  number={3},
  pages={524--538},
  year={1993},
  publisher={Optical Society of America}
}

@inproceedings{almudever2024,
  author    = {C. G. Almudever and R. Wille and 
               F. Sebastiano and N. Haider and 
               E. Alarcon},
  title     = {From Designing Quantum Processors 
               to Large-Scale Quantum Computing 
               Systems},
  booktitle = {Design, Automation and Test in 
               Europe (DATE)},
  year      = {2024}
}

@article{edaq2025,
  author  = {Y.-W. Zhao and others},
  title   = {{EDA-Q}: Electronic Design Automation 
             for Superconducting Quantum Chip},
  journal = {IEEE Transactions on Computer-Aided 
             Design of Integrated Circuits and 
             Systems},
  year    = {2025}
}

@article{leung2019deterministic,
  title={Deterministic bidirectional communication and remote entanglement generation between superconducting qubits},
  author={Leung, Nelson and Lu, Y and Chakram, S and Naik, RK and Earnest, N and Ma, R and Jacobs, K and Cleland, AN and Schuster, DI},
  journal={npj quantum information},
  volume={5},
  number={1},
  pages={18},
  year={2019},
  publisher={Nature Publishing Group UK London}
}








\end{document}